\documentclass[12pt,reqno]{amsart}
\usepackage{amssymb}
\usepackage{amsfonts}
\usepackage{indentfirst}
\usepackage{amsopn}
\usepackage{amsmath}
\usepackage{amsthm}
\usepackage{amscd}
\usepackage{graphicx}
\usepackage{epstopdf}

\makeatletter\@addtoreset {equation}{section}\makeatother

\begin{document}

\title[Instability of double-periodic waves]{\bf Instability of double-periodic waves\\ in the nonlinear Schr\"{o}dinger equation}

\author{Dmitry E. Pelinovsky}
\address{Department of Mathematics, McMaster University, Hamilton, Ontario, Canada, L8S 4K1}

\date{}
\maketitle

\begin{abstract}
	It is shown how to compute the instability rates for the double-periodic solutions to the cubic NLS (nonlinear Schr\"{o}dinger) equation by using the Lax linear equations. The wave function modulus of the  double-periodic solutions is periodic both in space and time coordinates; such solutions generalize the standing waves which have the time-independent and space-periodic wave function modulus. Similar to other waves in the NLS equation, the double-periodic solutions are spectrally unstable and this instability is related to the bands of the Lax spectrum outside the imaginary axis. A simple numerical method is used to compute the unstable spectrum and to compare the instability rates of the double-periodic solutions with those of the standing periodic waves.
\end{abstract}

\section{Introduction}

Peregrine breather is a rogue wave arising on the background of
the constant-amplitude wave due to its modulational instability \cite{Peregrine,Akh85}. The focusing cubic NLS (nonlinear Schr\"{o}dinger) equation is the canonical model which describes
both the modulational instability and the formation of rogue waves. 
Formation of rogue waves on the constant-amplitude background have been
modeled from different initial data such as local condensates \cite{Gelash1},
multi-soliton gases \cite{ZakGelash,Gelash2,GelashPRL}, and periodic perturbations \cite{AZ1,AZ2}.
Rogue waves have been experimentally observed both in hydrodynamical
and optical laboratories \cite{experiments1} (see recent reviews in \cite{Amin,Suret}).

Mathematical theory of rogue waves on the constant-amplitude background has seen 
many recent developments. 
Universal behavior of the modulationally unstable constant-amplitude background was studied asymptotically in
\cite{BiondiniPRL,Biondini}. The finite-gap method was employed to relate the unstable modes on the constant-amplitude background
with the occurrence of rogue waves \cite{GS1,GS2}. Rogue waves of infinite
order were constructed in \cite{Bil3} based on recent developments in the inverse scattering method \cite{Bil2}.
Rogue waves of the soliton superposition were studied asymptotically in the limit of many solitons \cite{PelSl,Bil1}.

At the same time, rogue waves were also investigated on the background of  standing periodic waves expressed by the Jacobian elliptic functions. Such exact solutions to the NLS equation were constructed first in \cite{CPnls} (see also early numerical work in \cite{Kedziora} and the recent generalization in \cite{Feng}). It was confirmed in \cite{CPW2} that 
these rogue waves arise due to the modulational instability of the standing periodic waves \cite{BHJ} (see also \cite{Kam,Kam-review}). Modulational instability of the periodic standing waves can be characterized by the separation of variables in the Lax system of linear equations \cite{DS} (see also \cite{DU,DU2}), compatibility of which gives the NLS equation. Modulational instability and rogue waves on the background of standing periodic waves have been experimentally observed in \cite{Xu}.

{\em The main goal of this paper is to compute the instability rates for the double-periodic solutions to the NLS equation, for which the wave function modulus is periodic with respect to both space and time coordinates.}  In particular, we consider two families of double-periodic solutions expressed as rational functions of the Jacobian elliptic functions which were constructed in the pioneering work \cite{Nail1}. These solutions represent perturbations of the Akhmediev breathers and describe generation of either phase-repeated or phase-alternating wave patterns 
\cite{Nail2,SciRep}. Rogue waves on the background of the double-periodic solutions were studied in \cite{CPW1} 
(see also numerical work in \cite{CalSch,Trillo}).  
Experimental observation of the double-periodic solutions in optical fibers was reported in \cite{TrilloOL}.

The double-periodic solutions constructed in \cite{Nail1} are particular cases of the quasi-periodic solutions of the NLS equation given by the Riemann Theta functions of genus two \cite{Smirnov1,Smirnov2,Wright}. Rogue waves for general quasi-periodic 
solutions of any genus were considered in \cite{Tovbis1,Tovbis2,Wright2}. 

Instability of the double-periodic solutions is studied using the Floquet theory for the Lax system of linear equations both in space and time coordinates.
We compute the instability rates of the double-periodic solutions and compare them with those for the standing periodic waves. In order to provide a fair comparison, we normalize the amplitude of all solutions to unity. {\em As a main ourcome of this work, we show that the instability rates are larger for the constant-amplitude waves and smaller for the double-periodic waves. }

The article is organized as follows. The explicit solutions to the NLS equation are reviewed in Section 2. Instability rates 
for the standing periodic waves and the double-periodic solutions are computed in Sections 3 and 4 respectively. Further directions are discussed in Section 5.

\section{Explicit solutions to the NLS equation}

The nonlinear Schr\"{o}dinger (NLS) equation is a fundamental model 
for nonlinear wave dynamics \cite{Sulem,Fibich}. 
We take the NLS equation in the standard form:
\begin{equation}
i \psi_t + \frac{1}{2} \psi_{xx} + |\psi|^2 \psi = 0.
\label{nls}
\end{equation}
This model has several physical symmetries which are checked directly:
\begin{itemize}
	\item {\em translation}
\begin{equation}
\label{translations}
\mbox{\rm if} \; \psi(x,t) \;\; \mbox{\rm is a solution, so is} \;\psi(x+x_0,t+t_0),
\;\; \mbox{\rm for every} \;\; (x_0,t_0) \in \mathbb{R} \times \mathbb{R},
\end{equation}
\item{\em scaling}
\begin{equation}
\label{scaling-transform}
\mbox{\rm if} \; \psi(x,t) \;\; \mbox{\rm is a solution, so is} \; \alpha \psi(\alpha x,\alpha^2 t),
\;\; \mbox{\rm for every} \;\; \alpha \in \mathbb{R},
\end{equation}
\item {\em Lorentz transformation:}
\begin{equation}
\label{Lorentz-transform}
\mbox{\rm if} \; \psi(x,t) \;\; \mbox{\rm is a solution, so is} \;\;  \psi(x + \beta t,t) e^{-i \beta x - \frac{i}{2} \beta^2 t},
\;\; \mbox{\rm for every} \;\; \beta \in \mathbb{R}.
\end{equation}
\end{itemize}
In what follows, we use the scaling symmetry (\ref{scaling-transform}) to normalize the amplitude of perodic and double-periodic solutions to unity and the Lorentz symmetry (\ref{Lorentz-transform}) to set the wave speed to zero. We also neglect the translational parameters $(x_0,t_0)$ due to the symmetry (\ref{translations}).

A solution $\psi(x,t) : \mathbb{R} \times \mathbb{R} \to \mathbb{C}$ to the NLS equation (\ref{nls}) is a compatibility
condition of the Lax system of linear equations on $\varphi(x,t) : \mathbb{R} \times \mathbb{R} \to \mathbb{C}^2$:
\begin{equation}\label{3.1}
\varphi_x = U(\lambda,\psi) \varphi,\qquad \qquad
U(\lambda,\psi) = \left(\begin{array}{cc} \lambda & \psi \\ -\bar{\psi} & -\lambda
\end{array} \right)
\end{equation}
and
\begin{equation}\label{3.2}
\varphi_t = V(\lambda,\psi) \varphi,\qquad
V(\lambda,\psi) = i \left(\begin{array}{cc}
\lambda^2 + \frac{1}{2} |\psi|^2 & \frac{1}{2} \psi_x + \lambda \psi\\
\frac{1}{2} \bar{\psi}_x - \lambda\bar{\psi} & -\lambda^2 - \frac{1}{2} |\psi|^2\\
\end{array}
\right),
\end{equation}
where $\bar{\psi}$ is the conjugate of $\psi$ and $\lambda \in \mathbb{C}$ is a spectral parameter.

The algebraic method developed in \cite{CPW1}
allows us to construct the stationary (Lax--Novikov) equations which admit a large class of bounded periodic and quasi-periodic solutions to the NLS equation (\ref{nls}). The simplest first-order Lax--Novikov equation is given by
\begin{equation}
\label{LN-1}
\frac{du}{dx} + 2 i c u = 0,
\end{equation}
where $c$ is arbitrary real parameter. A general
solution of this equation is given by $u(x) = A e^{-2ic x}$, where $A$ is the integration constant. This solution determines the  
constant-amplitude waves of the NLS equation (\ref{nls}) in the form:
\begin{equation}
\label{CW}
\psi(x,t) = A e^{-2ic (x+ct) + i A^2 t},
\end{equation}
where $A > 0$ is the constant amplitude and  
translations in $(x,t)$ are neglected due to the translational symmetry (\ref{translations}). 
Without loss of generality, $c$ can be set to $0$ due to the Lorentz transformation. Indeed, transformation 
(\ref{Lorentz-transform}) with $\beta = -c$ transforms 
(\ref{CW}) to the equivalent form $\psi(x,t) = A e^{-ic x - \frac{i}{2} c^2 t + i A^2 t}$, which is obtained from $\psi(x,t) = A e^{-i A^2 t}$ due to 
transformation (\ref{Lorentz-transform}) with $\beta = c$. By the scaling 
transformation (\ref{scaling-transform}) with $\alpha = A^{-1}$, the amplitude $A$ can be set to unity, which yields the normalized solution 
$\psi(x,t) = e^{it}$.

The second-order Lax--Novikov equation is given by 
\begin{equation}
\label{LN-2}
\frac{d^2 u}{dx^2} + 2 |u|^2 u + 2i c \frac{du}{dx} - 4 b u = 0,
\end{equation}
where $(c,b)$ are arbitrary real parameters. 
Solutions $u$ to the second-order equation (\ref{LN-2}) determines the standing travelling waves of the NLS equation (\ref{nls}) in the form:
\begin{equation}
\label{standing-wave}
\psi(x,t) = u(x+ct) e^{2ibt}.
\end{equation}
Without loss of generality, we set $c = 0$ due to 
the Lorentz transformation (\ref{Lorentz-transform}) with $\beta = -c$.  
Waves with the trivial phase are of particular interest \cite{CPnls,DS}. 
There are two families of such standing periodic waves given by the Jacobian elliptic functions in the form:
\begin{equation}
\label{red-1}
\psi(x,t) = {\rm dn}(x; k) e^{i (1-k^2/2) t} 
\end{equation}
and
\begin{equation}
\label{red-2}
\psi(x,t) = k {\rm cn}(x; k) e^{i (k^2 -1/2) t},
\end{equation}
where the parameter $k \in (0,1)$ is the elliptic modulus. The solutions 
(\ref{red-1}) and (\ref{red-2}) are defined up to the scaling transformation (\ref{scaling-transform}) and translations (\ref{translations}). The amplitude (maximal value of $|\psi|$) is 
set to unity for (\ref{red-1}) and to $k$ for (\ref{red-2}). In order to normalize 
the amplitude to unity for the cnoidal wave (\ref{red-2}),  we can use the scaling transformation (\ref{scaling-transform}) with 
$\alpha = k^{-1}$.

Due to the well-known expansion formulas 
\begin{eqnarray*}
{\rm dn}(x;k) & = & {\rm sech}(x) + \frac{1}{4} (1-k^2) \left[ \sinh(x) \cosh(x) + x \right] \tanh(x) {\rm sech}(x) + \mathcal{O}((1-k^2)^2), \\
{\rm cn}(x;k) & = & {\rm sech}(x) - \frac{1}{4} (1-k^2) \left[ \sinh(x) \cosh(x) - x \right] \tanh(x) {\rm sech}(x) + \mathcal{O}((1-k^2)^2),
\end{eqnarray*}
both the periodic waves (\ref{red-1}) and (\ref{red-2}) approaches 
the NLS soliton $\psi(x,t) = {\rm sech}(x) e^{i t/2}$ as $k \to 1$. In the other limit, 
the dnoidal periodic wave (\ref{red-1}) approaches the constant-amplitude 
wave $\psi(x,t) = e^{it}$ as $k \to 0$, whereas the normalized cnoidal periodic wave (\ref{red-2}) 
approaches the constant-amplitude wave $\psi(x,t) \sim \cos(x/k) e^{-it/(2k^2)}$ as $k \to 0$.

The third-order Lax--Novikov equation is given by 
\begin{equation}
\label{LN-3}
\frac{d^3 u}{dx^3} + 6 |u|^2 \frac{du}{dx} + 2i c 
\left( \frac{d^2 u}{dx^2} + 2|u|^2 u \right) - 4 b \frac{du}{dx} + 8i a u = 0,
\end{equation}
where $(a,b,c)$ are arbitrary real parameters. 
Waves with $a = c = 0$ are again of particular interest \cite{Nail1}. After a transformation of variables \cite{CPW1}, such solutions can be written in the form:
\begin{equation}
\label{double-periodic}
\psi(x,t) = \left[ Q(x,t) + i \delta(t) \right] e^{i \theta(t)},
\end{equation}
where $Q(x,t) : \mathbb{R}\times \mathbb{R}\to \mathbb{R}$ is periodic both in the space and time coordinates and $\delta(t) : \mathbb{R}\to \mathbb{R}$ 
has a double period in $t$ compared to $Q(x,t)$. There are two particular families 
of the double-periodic solutions (\ref{double-periodic}), which can be written by using the Jacobian elliptic functions (see Appendices A and B in \cite{CPW1}):
\begin{equation}
\label{solB}
\psi(x,t) = k \frac{{\rm cn}(t;k) {\rm cn}(\sqrt{1+k}x;\kappa) + i \sqrt{1+k}
	{\rm sn}(t;k) {\rm dn}(\sqrt{1+k} x; \kappa)}{\sqrt{1+k} {\rm
		dn}(\sqrt{1+k}x;\kappa) - {\rm dn}(t;k) {\rm cn}(\sqrt{1+k} x; \kappa)} e^{i
	t}, \quad \kappa = \frac{\sqrt{1-k}}{\sqrt{1+k}}
\end{equation}
and
\begin{equation}
\label{solA}
\psi(x,t) = \frac{{\rm dn}(t;k) {\rm cn}(\sqrt{2} x; \kappa) + i \sqrt{k(1+k)} {\rm sn}(t;k)}{\sqrt{1+k}
	- \sqrt{k} {\rm cn}(t;k) {\rm cn}(\sqrt{2} x; \kappa)} e^{i k t}, \quad \kappa
= \frac{\sqrt{1-k}}{\sqrt{2}},
\end{equation}
where $k \in (0,1)$ is the elliptic modulus. The solutions (\ref{solB}) 
and (\ref{solA}) are defined up to the scaling transformation (\ref{scaling-transform}) and the translations (\ref{translations}). 
The amplitude (maximal value of $|u|$) is 
$\sqrt{1+k} + 1$ for (\ref{solB}) and $\sqrt{1+k} + \sqrt{k}$ for (\ref{solA}). In order to normalize 
the amplitudes of the double-periodic waves to unity,  we can use the scaling transformation (\ref{scaling-transform}) with 
$\alpha = (\sqrt{1+k} + 1)^{-1}$ and $\alpha = (\sqrt{1+k} + \sqrt{k})^{-1}$ respectively.

The double-periodic solutions (\ref{solB}) and (\ref{solA}) can be written in the form:
\begin{equation}
\label{double-periodic-mod}
\psi(x,t) = \phi(x,t) e^{2i b t}, \qquad \phi(x+L,t) = \phi(x,t+T) = \phi(x,t),
\end{equation}
where $L > 0$ and $T > 0$ are fundamental periods in space and time coordinates, respectively, whereas $2b = 1$ for (\ref{solB}) 
and $2b = k$ for (\ref{solA}). 

\begin{figure}[htp!]
	\centering
	\includegraphics[width=8cm,height=5cm]{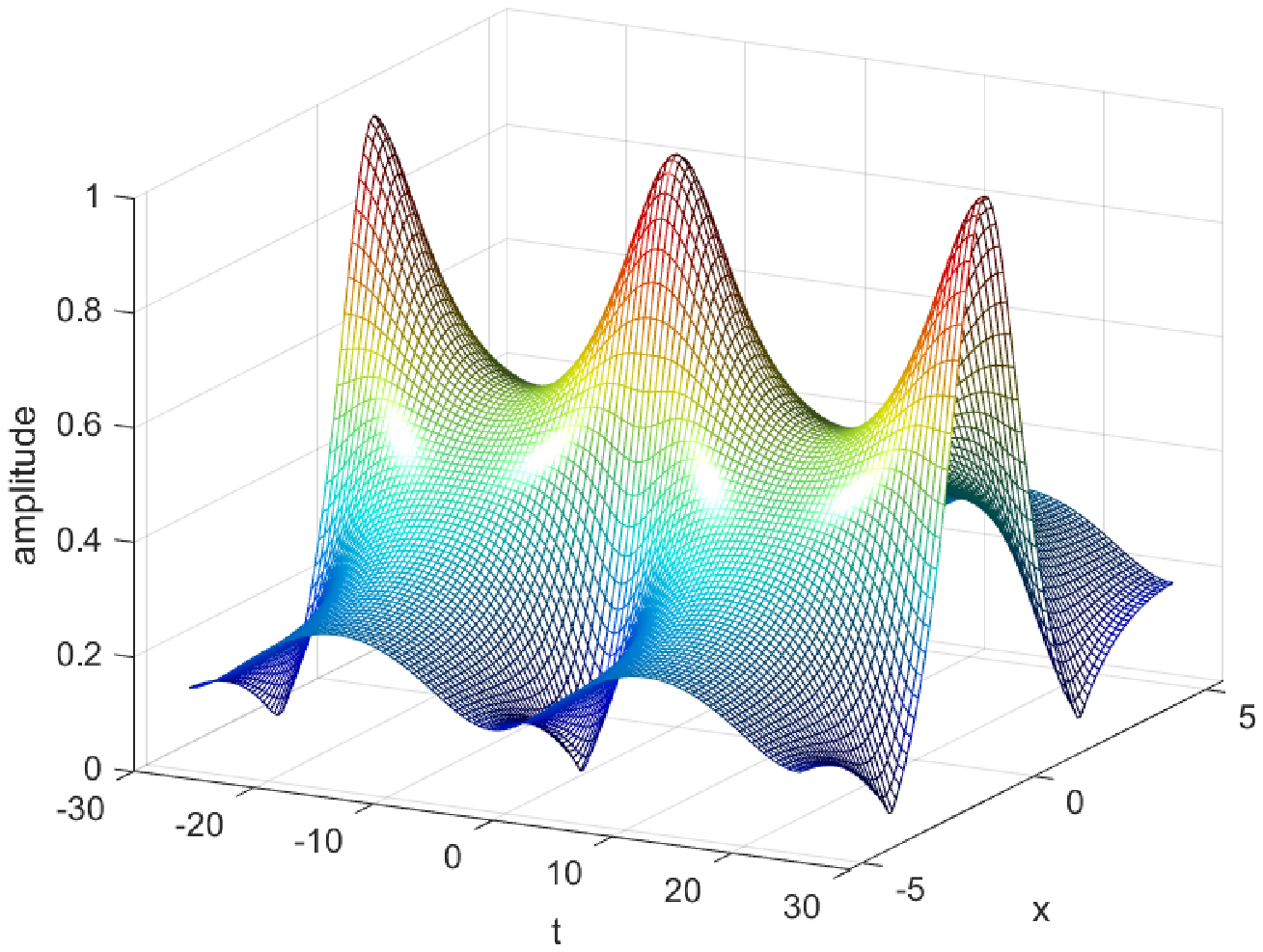}
	\includegraphics[width=8cm,height=5cm]{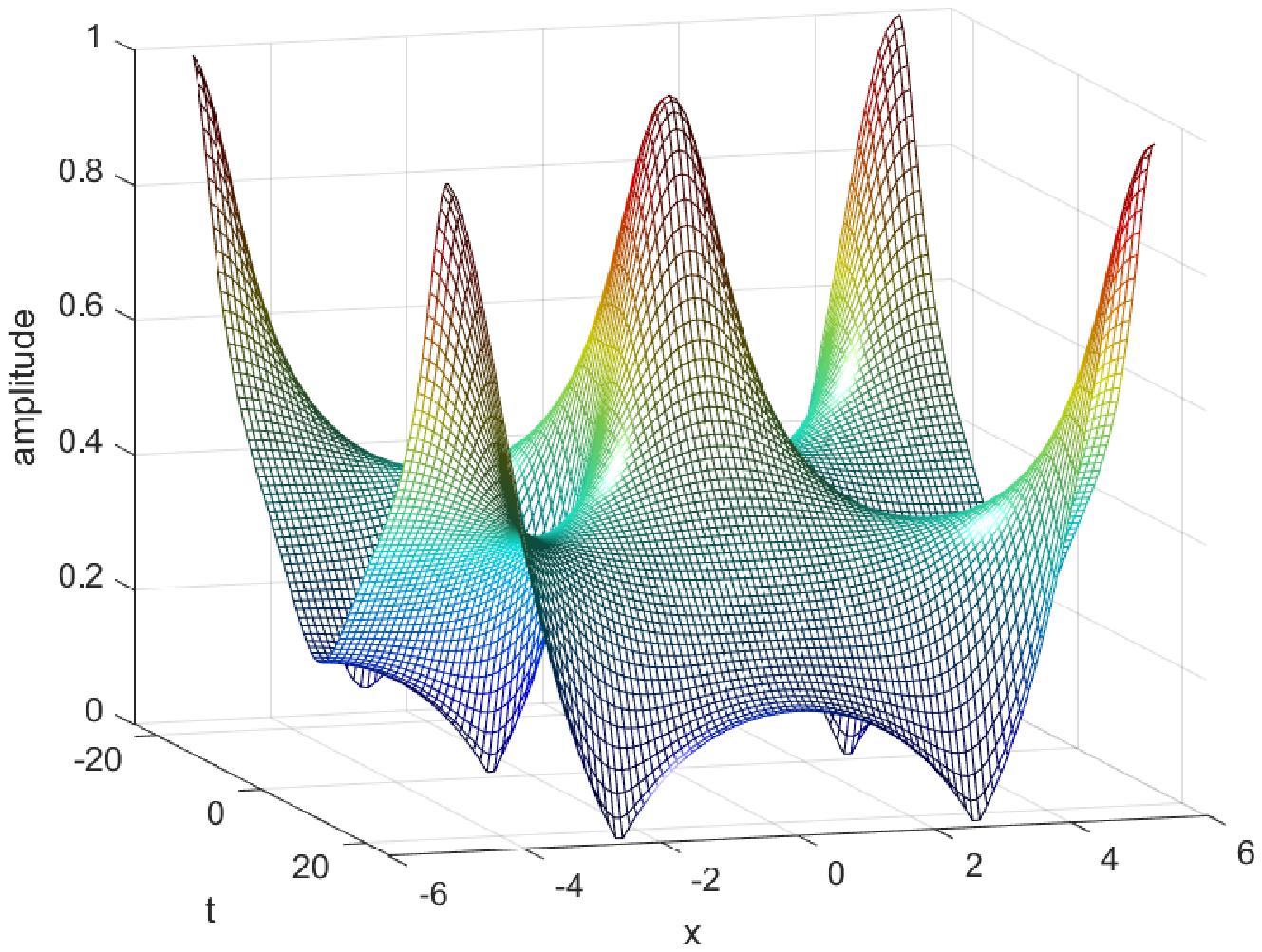}
	\caption{Amplitude-normalized double-periodic waves (\ref{solB}) (left) 
		and (\ref{solA}) (right) with $k = 0.9$.}
	\label{f00}
\end{figure}

Figure \ref{f00} shows surface plots of $|\psi|$ 
on the $(x,t)$ plane within the fundamental periods. 
The amplitudes of the double-periodic waves on Fig. \ref{f00} have been normalized to unity by the scaling transformation (\ref{scaling-transform}).
The solution (\ref{solB}) generates the phase-repeated wave patterns, whereas 
the solution (\ref{solA}) generates the phase-alternating patterns 
\cite{Nail2,SciRep,CPW1,Trillo}.

As $k \to 1$, both the double-periodic solutions (\ref{solB}) and (\ref{solA}) approach to the same Akhmediev breather given by 
\begin{equation}
\label{Akh-breather}
\psi(x,t) = \frac{\cos(\sqrt{2} x) + i \sqrt{2} \sinh(t)}{\sqrt{2} \cosh(t) - \cos(\sqrt{2} t)} e^{it}.
\end{equation}
As $k \to 0$, the solution (\ref{solB}) approaches the scaled NLS soliton 
\begin{equation}
\label{NLS-soliton}
\psi(x,t) = 2 {\rm sech}(2x) e^{2it},
\end{equation} 
whereas the solution (\ref{solA}) approaches the scaled cnoidal wave 
\begin{equation}
\label{NLS-wave}
\psi(x,t) = {\rm cn}\left(\sqrt{2}x;\frac{1}{\sqrt{2}}\right).
\end{equation}
These limits are useful to control accuracy of numerical 
computations of the modulational instability rate 
for the double-periodic solutions in comparison with 
the similar numerical computations for the standing waves.

\section{Instability of standing waves}

Here we review how to use the linear equations (\ref{3.1})--(\ref{3.2}) 
in order to compute the instability rates for the standing periodic waves  (\ref{standing-wave}) (see \cite{CPW2,DS}).
Due to the separation of variables in (\ref{standing-wave}), one can write
\begin{equation}
\label{lax-eq}
\varphi_1(x,t) = \chi_1(x+ct) e^{ibt + t \Omega}, \quad
\varphi_2(x,t) = \chi_2(x+ct) e^{-ibt + t \Omega},
\end{equation}
where $\Omega \in \mathbb{C}$ is another spectral parameter 
and $\chi = (\chi_1,\chi_2)^T$ satisfies the following spectral problems:
\begin{equation}
\label{lin-alg-1}
\chi_x = \left(\begin{array}{cc} \lambda & u \\ -\bar{u} & -\lambda \end{array} \right) \chi, 
\end{equation}
and
\begin{equation}
\label{lin-alg-2}
\Omega \chi + c \left(\begin{array}{cc} \lambda & u \\ -\bar{u} & -\lambda \end{array} \right) \chi = i \left(\begin{array}{cc}
\lambda^2 + \frac{1}{2} |u|^2 - b & \frac{1}{2} \frac{du}{dx} + \lambda u \\
\frac{1}{2} \frac{d \bar{u}}{dx} - \lambda\bar{u} & -\lambda^2 - \frac{1}{2} |u|^2 + b \end{array} \right) \chi.
\end{equation}

We say that $\lambda$ belongs to {\em the Lax spectrum} of the spectral problem 
(\ref{lin-alg-1}) if $\chi \in L^{\infty}(\mathbb{R})$. 
Since $u(x+L) = u(x)$ is periodic with the fundamental period $L > 0$, 
Floquet's Theorem guarantees that bounded solutions
of the linear equation (\ref{lin-alg-1}) can be represented in the form:
\begin{equation}\label{A1FB}
\chi(x) = \hat{\chi}(x) e^{i \theta x},
\end{equation}
where $\hat{\chi}(x+L) = \hat{\chi}(x)$ and $\theta \in \left[-\frac{\pi}{L},\frac{\pi}{L} \right]$. When $\theta = 0$ and 
$\theta = \pm \frac{\pi}{L}$, the bounded solutions (\ref{A1FB}) are periodic and anti-periodic, respectively.

Since the spectral problem (\ref{lin-alg-2}) is a linear algebraic system, it admits a nonzero solution if and only if
the determinant of the coefficient matrix is zero. The latter condition yields
the $x$-independent relation between $\Omega$ and $\lambda$ in the form
$\Omega^2 + P(\lambda) = 0$, where $P(\lambda)$ is given by 
\begin{equation}
\label{Polynomial-2}
P(\lambda) = \lambda^4 + 2 i c \lambda^3 - (c^2 + 2b) \lambda^2 + 2i (a - bc) \lambda + b^2 - 2 ac + 2 d,
\end{equation}
with parameters $a$ and $d$ being the conserved quantities of the second-order equation (\ref{LN-2}):
\begin{equation}
\label{Constant-2b}
\left|  \frac{du}{dx} \right|^2 + |u|^4 - 4 b |u|^2 = 8 d
\end{equation}
and
\begin{equation}
\label{Constant-2a}
i \left( \frac{du}{dx}\bar{u} - u  \frac{d \bar{u}}{dx} \right) - 2 c |u|^2 = 4 a.
\end{equation}
Polynomial $P(\lambda)$ naturally occurs in the algebraic method \cite{CPW1}. For the standing waves of the trivial phase with $a = c = 0$, 
the polynomial $P(\lambda)$ can be written explicitly in the form:
\begin{equation}
\label{P-poly-0}
P(\lambda) = \lambda^4 - \frac{1}{2} (u_1^2 + u_2^2) \lambda^2 + \frac{1}{16} (u_1^2 - u_2^2)^2,
\end{equation}
where the turning points $u_1$ and $u_2$ parameterize $b$ and $d$ in the form:
\begin{equation}
\label{roots-0-connection}
\left\{ \begin{array}{l}
4b = u_1^2 + u_2^2, \\
8d = -u_1^2 u_2^2.
\end{array} \right.
\end{equation}
Roots of $P(\lambda)$ are located at $\{ \pm \lambda_1, \pm \lambda_2\}$ given by 
\begin{equation}
\label{eig-dn-0}
\lambda_1 = \frac{u_1 + u_2}{2}, \quad
\lambda_2 = \frac{u_1 - u_2}{2},
\end{equation}
so that the polynomial $P(\lambda)$ can be written in the factorized form:
\begin{equation}
\label{factorization}
P(\lambda) = (\lambda^2 - \lambda_1^2) (\lambda^2 - \lambda_2^2).
\end{equation}

By adding a perturbation $v$ to the standing wave $u$ in the form
\begin{equation}
\psi(x,t) = e^{2ibt} \left[ u(x+ct) + v(x+ct,t) \right]
\end{equation}
and dropping the quadratic terms in $v$, we obtain 
the linearized system of equations which describe linear stability 
of the standing waves (\ref{standing-wave}):
\begin{equation}
\left\{ \begin{array}{l}
i v_t - 2 b v + i c v_x + \frac{1}{2} v_{xx} + 2|u|^2 v + u^2 \bar{v} = 0, \\
-i \bar{v}_t - 2 b \bar{v} - i c \bar{v}_x + \frac{1}{2} \bar{v}_{xx}
+ 2|u|^2 \bar{v} + \bar{u}^2 v = 0.
\end{array} \right.
\label{nls-lin}
\end{equation}
The variables can be separated in the form:
\begin{equation}
\label{w-eigen}
v(x,t) = w_1(x) e^{t \Lambda}, \quad \bar{v}(x,t) = w_2(x) e^{t \Lambda},
\end{equation}
where $\Lambda$ is a spectral parameter and $w = (w_1,w_2)^T$ satisfies the spectral stability problem 
\begin{equation}
i \Lambda \sigma_3 w  + \left( \begin{array}{cc} 
\frac{1}{2} \partial_x^2 + 2|u|^2 - 2b + i c \partial_x  & u^2 \\
\bar{u}^2 & \frac{1}{2} \partial_x^2 + 2|u|^2 - 2b - i c \partial_x 
\end{array} \right) w = 0,
\label{nls-spectral-stab}
\end{equation}
where $\sigma_3 = {\rm diag}(1,-1)$.
Note that $w_1$ and $w_2$ are no longer complex conjugate 
if $\Lambda \notin \mathbb{R}$. 

We say that $\Lambda$ belongs to {\em the stability spectrum} of the spectral problem (\ref{nls-spectral-stab}) if $w \in L^{\infty}(\mathbb{R})$. If $\lambda$ is in the Lax spectrum of the spectral problem (\ref{lin-alg-1}), then the bounded 
squared eigenfunctions $\chi_1^2$ and $\chi_2^2$ determine 
the bounded eigenfunctions $w_1$ and $w_2$ of the spectral stability problem (\ref{nls-spectral-stab}) and $\Omega$ determines eigenvalues $\Lambda$ as follows:
\begin{equation}
w_1 = \chi_1^2, \quad w_2 = -\chi_2^2, \quad 
\Lambda = 2 \Omega.
\label{eig-stability}
\end{equation}
Validity of (\ref{eig-stability}) can be checked directly from 
(\ref{lin-alg-1}), (\ref{lin-alg-2}), (\ref{w-eigen}), and (\ref{nls-spectral-stab}). If ${\rm Re}(\Lambda) > 0$ for $\lambda$ in the Lax spectrum, the periodic standing wave (\ref{standing-wave}) is
called {\em spectrally unstable}. It is called {\em modulationally unstable}
if the unstable spectrum with ${\rm Re}(\Lambda) > 0$ intersects the origin in the $\Lambda$-plane transversely to the imaginary axis. 

The importance of distinguishing between spectral and modulational instability of the periodic standing waves appears in the existence of rogue waves 
on their background. It was shown in \cite{PelWright} that 
if the periodic standing waves are spectrally unstable but modulationally stable, the rogue waves are not fully localized and degenerate into propagating algebraic solitons. Similarly, it was shown in \cite{CPW2} that 
if the unstable spectrum with ${\rm Re}(\Lambda) > 0$ intersects 
the origin in the $\Lambda$-plane tangentially to the imaginary axis, 
the corresponding rogue wave degenerates into a propagating algebraic soliton. 

Next, we compute the instability rates for the standing periodic waves (\ref{standing-wave}) of the trivial phase with $a = c = 0$.
It follows from (\ref{factorization}) with either real $\lambda_1$, $\lambda_2$ or complex-conjugate $\lambda_1 = \bar{\lambda}_2$ that if $\lambda \in i \mathbb{R}$ belongs to the Lax spectrum, then $\Lambda \in i \mathbb{R}$ belongs to the stable spectrum. Thus, the spectral instability of the standing periodic 
waves of the trivial phase is only related to the Lax spectrum with $\lambda \notin i \mathbb{R}$.

\begin{figure}[htpb!]
	\centering
	\includegraphics[width=8cm,height=5cm]{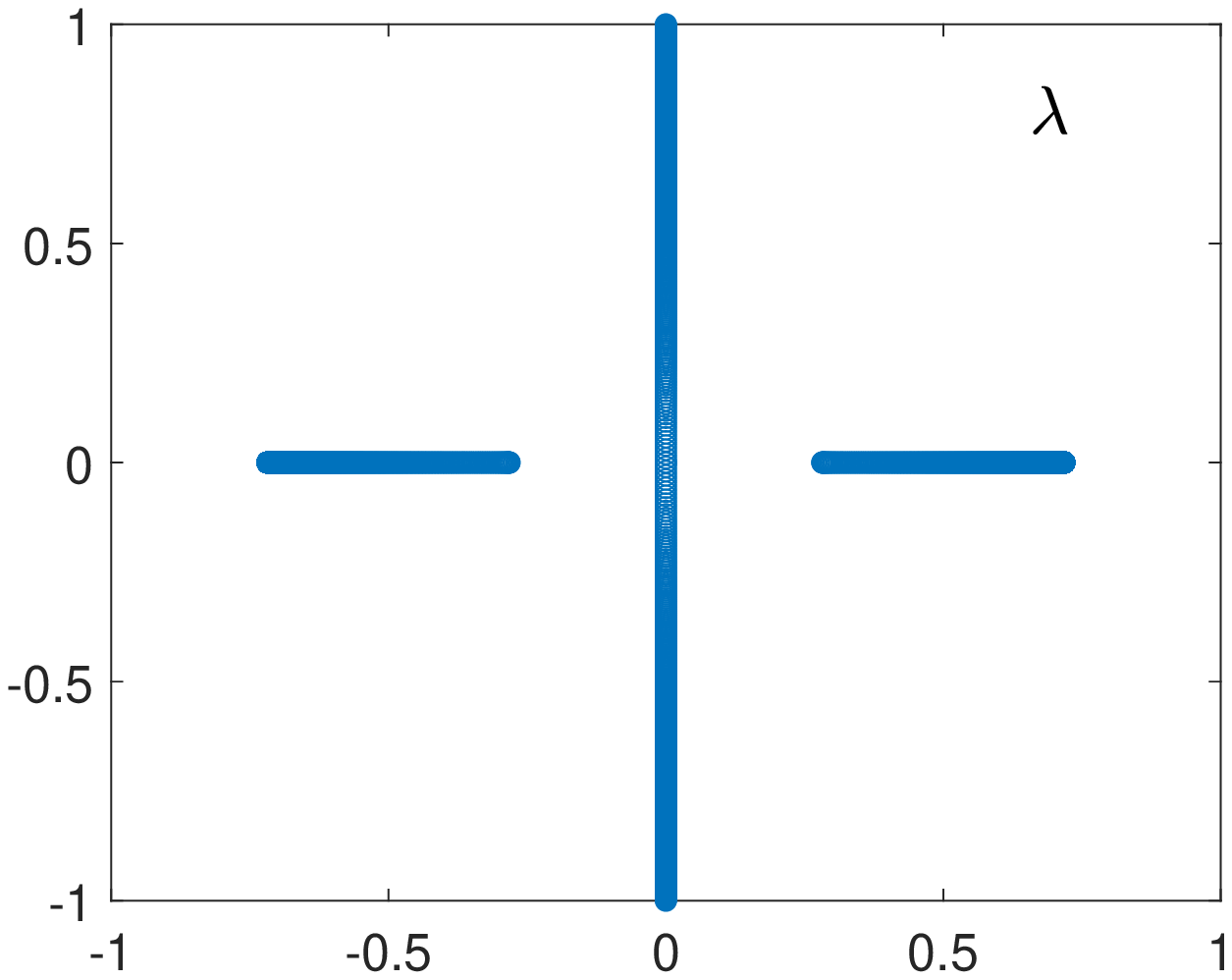}
	\includegraphics[width=8cm,height=5cm]{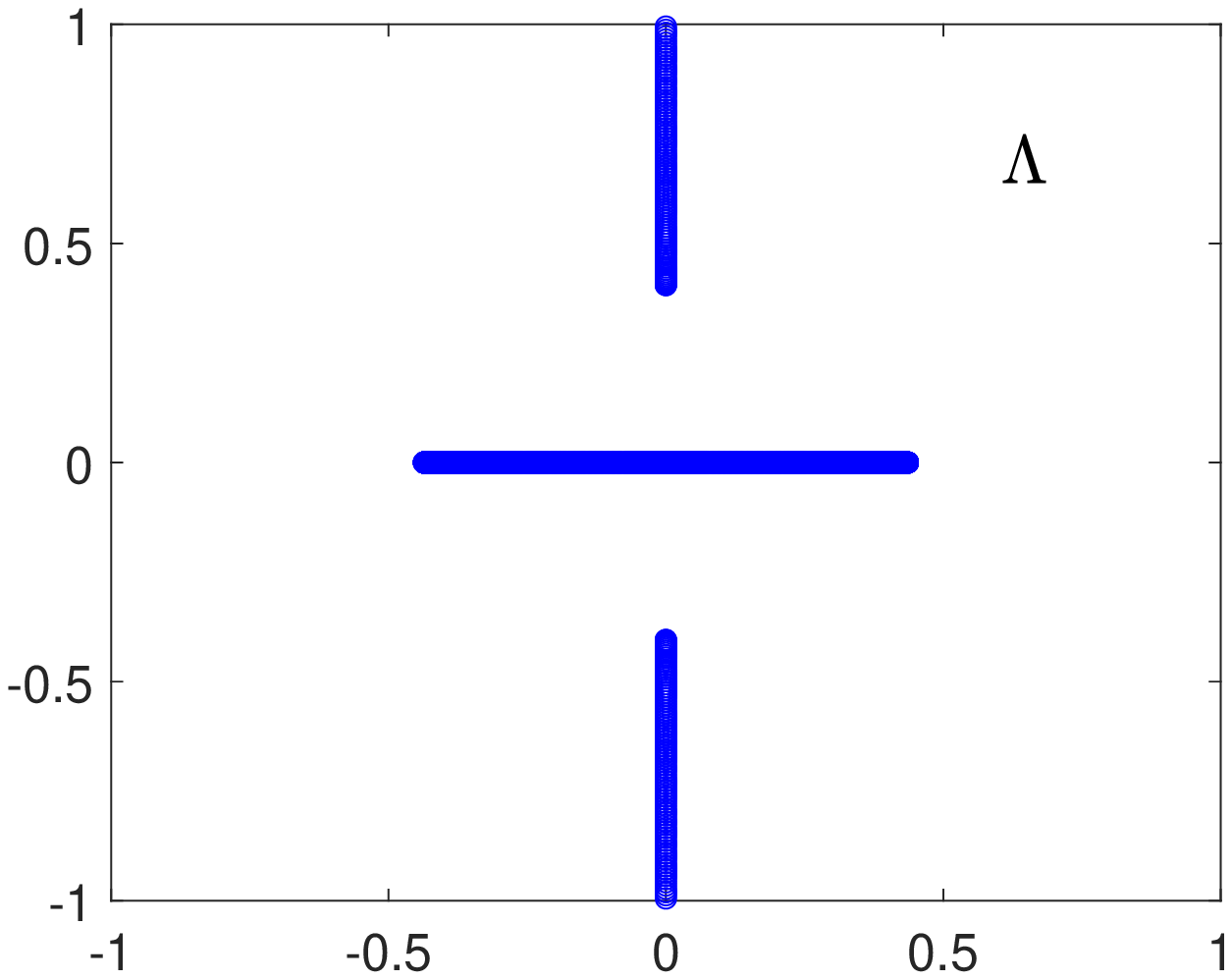}
	\caption{Lax spectrum on the $\lambda$-plane (left) and stability spectrum on the $\Lambda$-plane (right) for the ${\rm dn}$-periodic wave (\ref{red-1}) with $k=0.9$.}
	\label{f1}
\end{figure}

For the ${\rm dn}$-periodic wave (\ref{red-1}) with $u_1 = 1$ and $u_2 = \sqrt{1-k^2}$, the amplitude is already normalized to unity and no scaling transformation is needed. Lax spectrum of the spectral problem (\ref{lin-alg-1}) is shown on Fig. \ref{f1} (left) for $k = 0.9$. It follows from (\ref{factorization}) that $P(\lambda) < 0$ for $\lambda \in (\lambda_2,\lambda_1)$ with $P(\lambda_{1,2}) = 0$, where $\lambda_{1,2}$ are given by (\ref{eig-dn-0}). 
The unstable spectrum on the $\Lambda$-plane 
belongs to the finite segment on the real line which touches the origin as is shown on Fig. \ref{f1} (right), hence the ${\rm dn}$-periodic wave (\ref{red-1}) is both spectrally and modulationally unstable. It follows from (\ref{eig-stability}) that 
$$
\max_{\lambda \in [\lambda_2,\lambda_1]} \Lambda = \max_{\lambda \in [\lambda_2,\lambda_1]} 2 \sqrt{(\lambda_1^2 - \lambda^2)(\lambda^2 - \lambda_2^2)} = (\lambda_1^2 - \lambda_2^2) = \sqrt{1-k^2}.
$$
Since the ${\rm dn}$-periodic wave becomes the constant-amplitude wave of unit amplitude if $k = 0$, it is clear that the maximal instability rate is largest for the constant-amplitude wave with $k = 0$, monotonically decreasing in $k$, and vanishes for the soliton limit $k = 1$. As $k \to 1$, the horizontal band on Fig. \ref{f1} (right) shrinks to an eigenvalue at the origin.

\begin{figure}[htpb!]
	\centering
	\includegraphics[width=8cm,height=5cm]{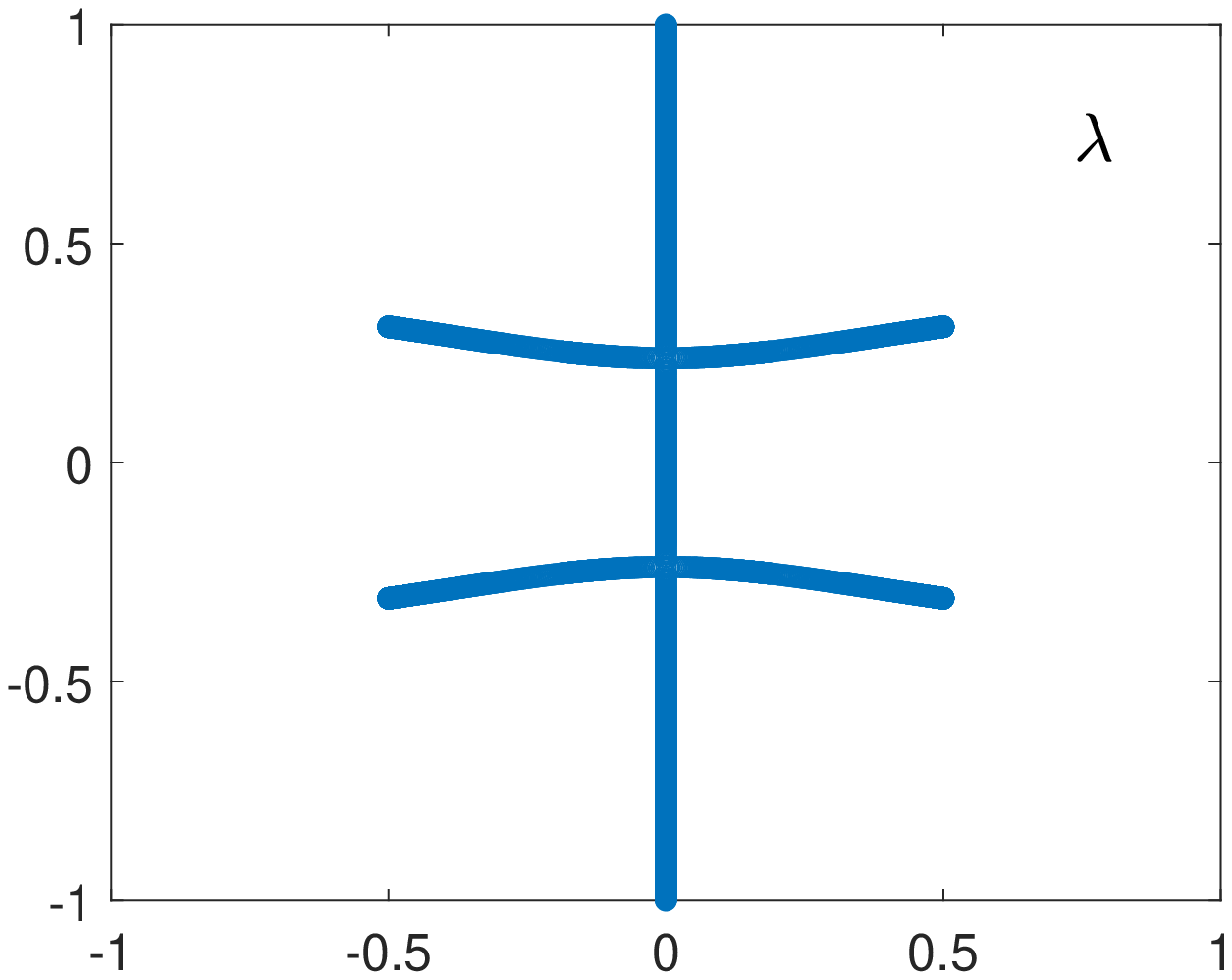}
	\includegraphics[width=8cm,height=5cm]{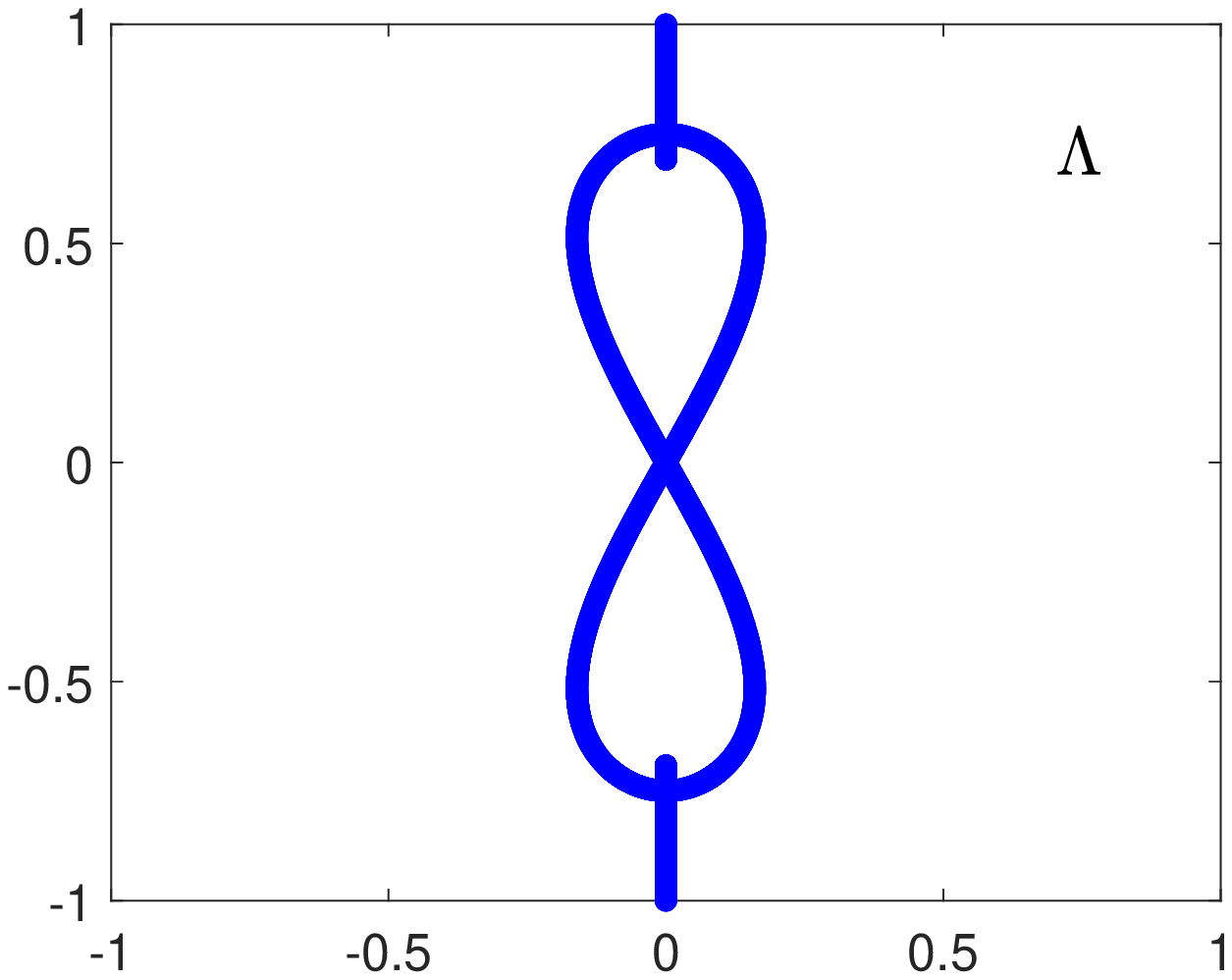}\\
	\includegraphics[width=8cm,height=5cm]{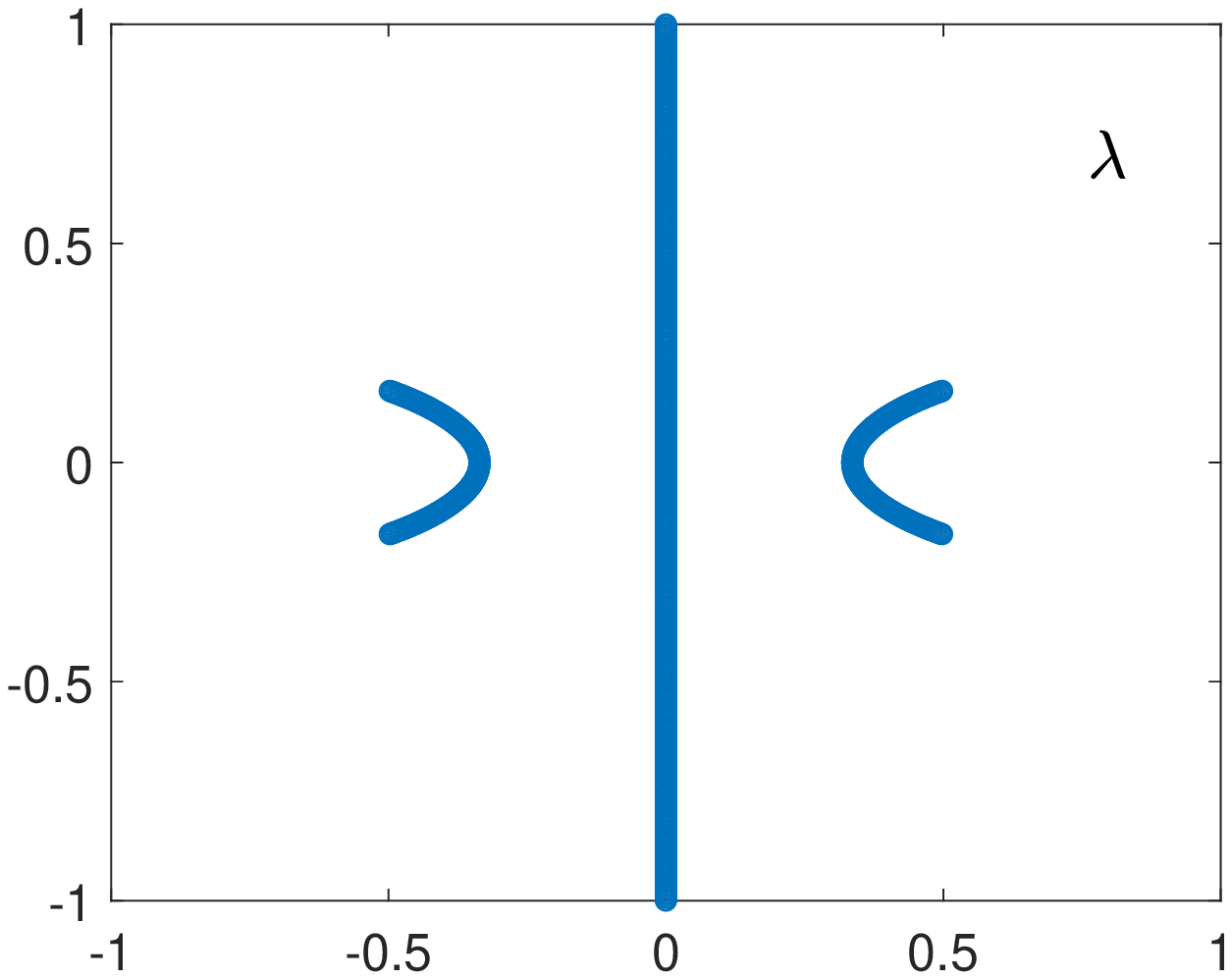} 
	\includegraphics[width=8cm,height=5cm]{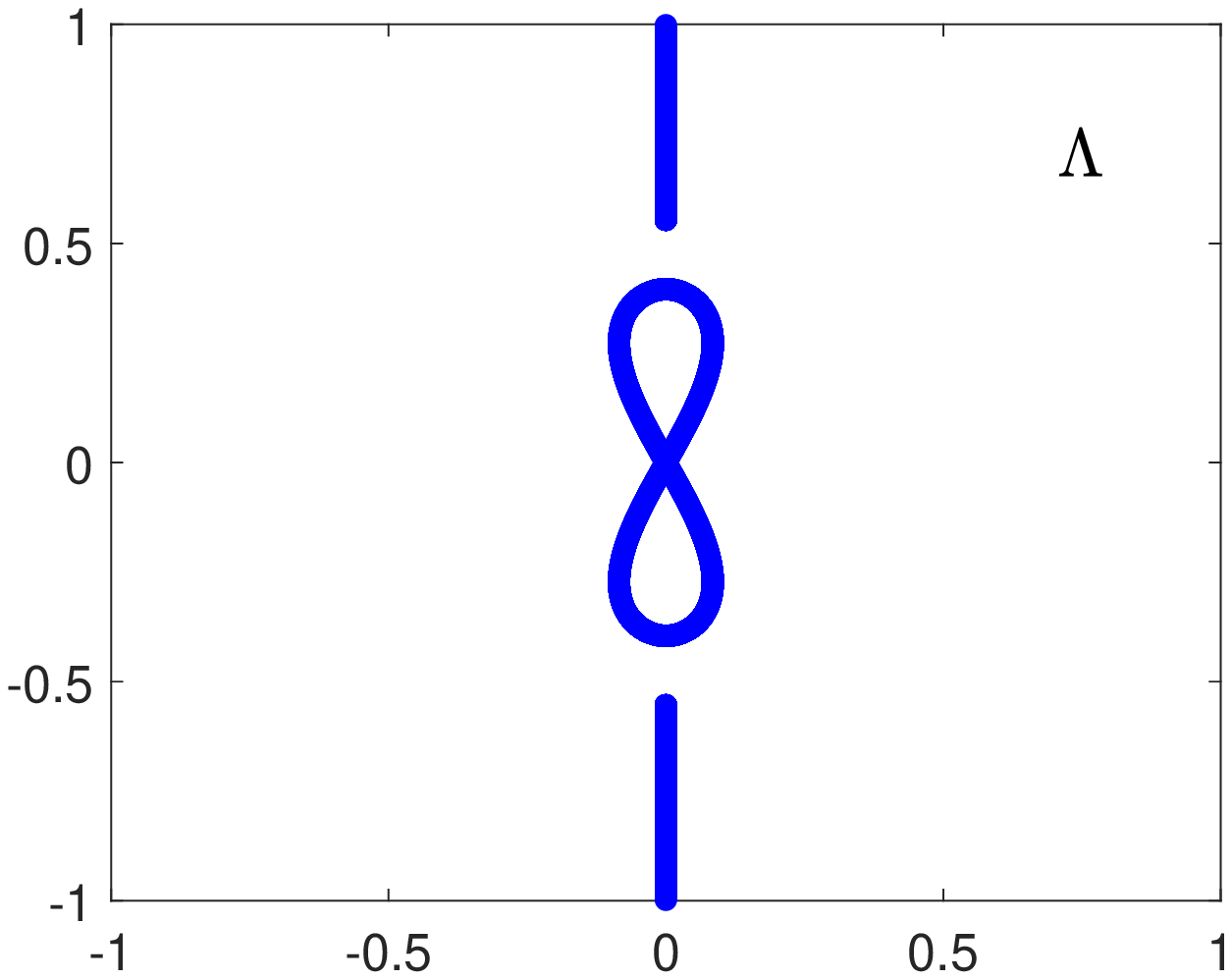} \\
	\caption{The same as Figure \ref{f1} but for the amplitude-normalized ${\rm cn}$-periodic wave with $k=0.85$ (top) and $k=0.95$ (bottom). }
	\label{f2}
\end{figure}

For the ${\rm cn}$-periodic wave (\ref{red-2}) with $u_1 = k$ and $u_2 = i \sqrt{1-k^2}$, the amplitude is $k$. Hence, we use the scaling transformation (\ref{scaling-transform}) with $\alpha = k^{-1}$ in order to normalize the amplitude to unity. Lax spectrum of the spectral problem (\ref{lin-alg-1}) 
for such an amplitude-normalized ${\rm cn}$-periodic wave is shown on left panels of Fig. \ref{f2} for $k = 0.85$ (top) and $k = 0.95$ (bottom). The unstable spectrum on the $\Lambda$-plane is obtained from the same expressions $\Lambda = \pm 2 i \sqrt{P(\lambda)}$ when $\lambda$ traverses along the bands of the Lax spectrum outside $i \mathbb{R}$. The unstable spectrum resembles the figure-eight band as is shown on the right panels of Fig. \ref{f2}. The figure-eight band starts and ends at $\Omega = 0$ for $\lambda = \lambda_1$ and $\lambda = \lambda_2$. Stability spectrum for both examples is similar in spite of the differences in the Lax spectrum. The only difference is that the figure-eight band and the purely imaginary bands intersect for $k = 0.85$ (top) and do not intersect for $k = 0.95$ (bottom). Thus, the ${\rm cn}$-periodic wave (\ref{red-2}) is spectrally and modulationally unstable. As $k \to 1$, the figure-eight band shrinks to an eigenvalue at the origin.

\begin{figure}[htp!]
	\centering
	\includegraphics[width=8cm,height=5cm]{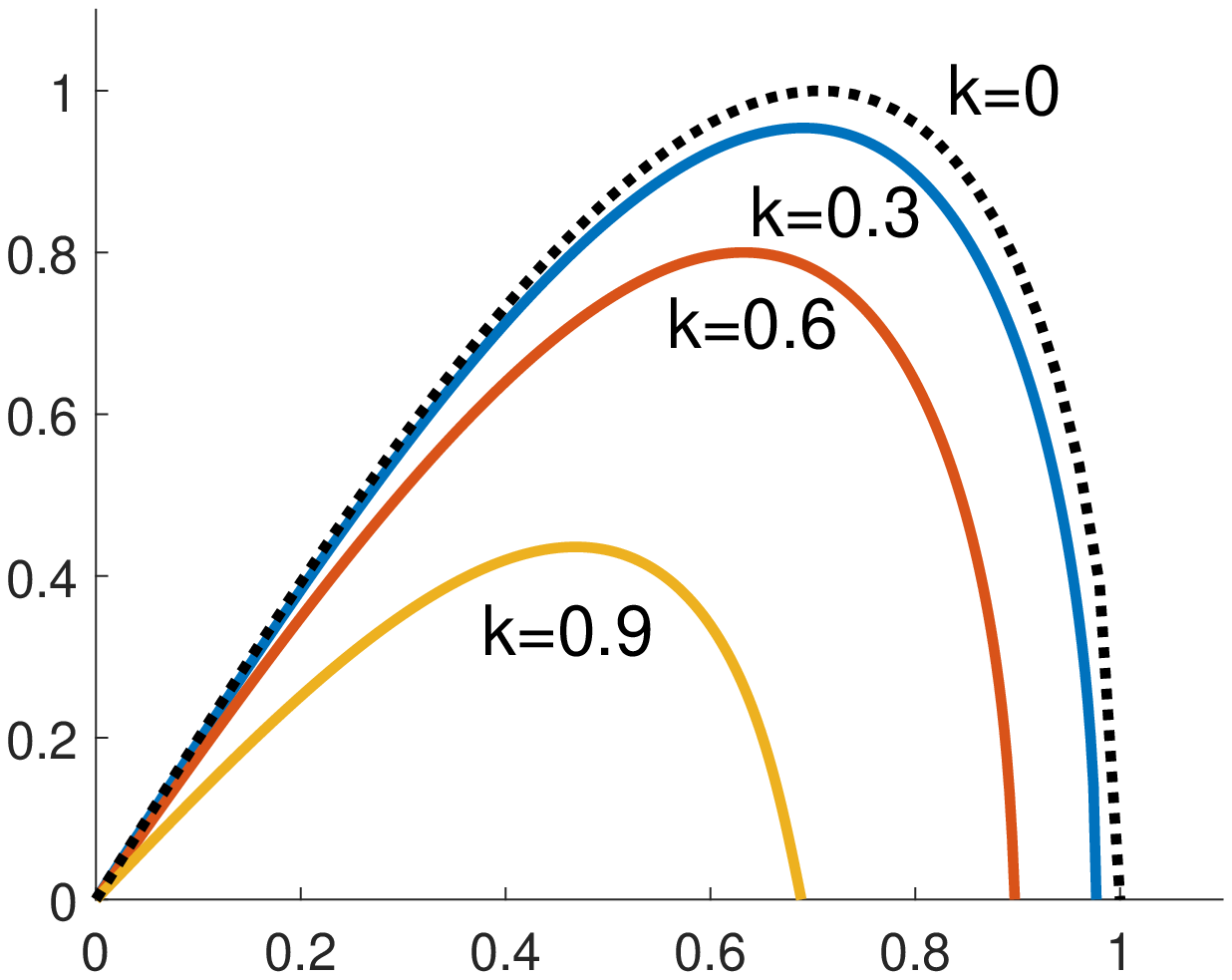}
	\includegraphics[width=8cm,height=5cm]{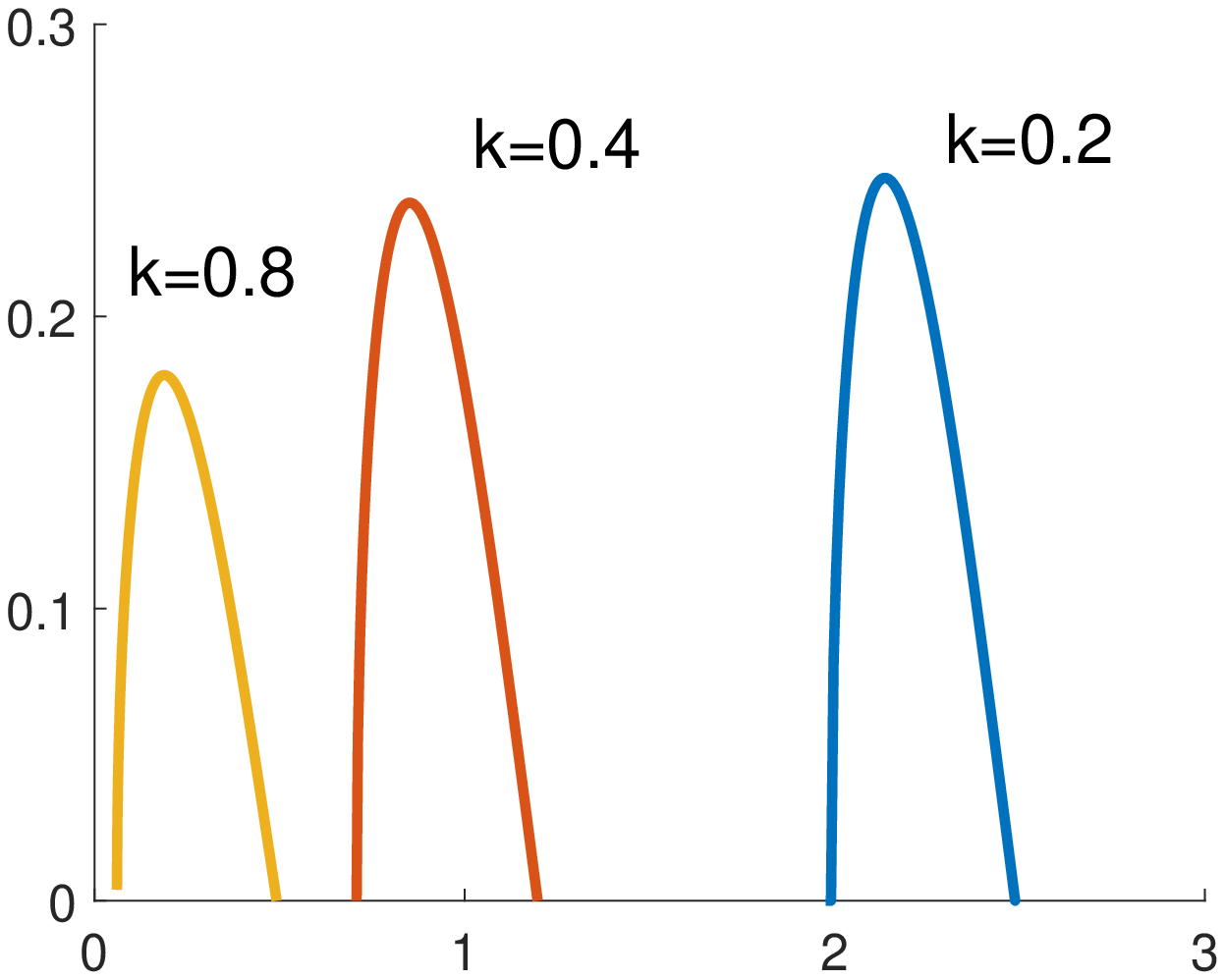}
	\caption{Instability rate ${\rm Re}(\Lambda)$ versus the Floquet parameter $\theta$ for the amplitude-normalized ${\rm dn}$-periodic (left) and ${\rm cn}$-periodic (right) waves. The values of $k$ in the elliptic functions are given in the plots.}
	\label{f0}
\end{figure}

Figure \ref{f0} compares the instability rates for different standing waves of the same unit amplitude. ${\rm Re}(\Lambda)$ is plotted versus the Floquet parameter $\theta$ in $[0,\frac{\pi}{L}]$ in (\ref{A1FB}).
For the ${\rm dn}$-periodic wave (left), we confirm that the growth rate is maximal for the constant-amplitude wave $(k=0)$ 
and is monotonically decreasing as $k$ is increased in $(0,1)$. 
For the ${\rm cn}$-periodic wave (right), the growth rate is also maximal in the limit $k \to 0$, for which the amplitude-normalized ${\rm cn}$-periodic wave is expanded as
\begin{equation}
\label{cn-expansion}
u(x) = {\rm cn}(k^{-1} x;k) \sim 0.5 e^{i k^{-1} x} + 0.5 e^{-i k^{-1} x} \quad \mbox{\rm as } \quad k \to 0. 
\end{equation}
Due to the scaling transformation (\ref{scaling-transform}) and the expansion (\ref{cn-expansion}), the maximal growth rate in the limit $k \to 0$ is $0.25$ instead of $1$ 
and the Floquet parameter $\theta$, for which it is attained, diverges to 
infinity as $k \to 0$. As $k$ increases in $(0,1)$, the growth rate becomes smaller and the Floquet parameter $\theta$ for which ${\rm Re}(\Lambda) > 0$ 
moves towards the origin. The end point of the unstable band reaches $\theta = 0$, when the bands of the Lax spectrum outside $i \mathbb{R}$ do not intersect $i \mathbb{R}$ like on Fig. \ref{f2} (bottom).

\section{Instability of double-periodic waves}

Here we describe the main result on how to compute the instability 
rates for the double-periodic waves by using the linear equations (\ref{3.1})--(\ref{3.2}). We write the solutions (\ref{solB}) and (\ref{solA}) in the form (\ref{double-periodic-mod}). We represent solution $\varphi$ to the linear equations (\ref{3.1})--(\ref{3.2}) in the form:
\begin{equation}
\label{lax-eq-mod}
\varphi_1(x,t) = \chi_1(x,t) e^{ib t + x \mu + t \Omega}, \quad
\varphi_2(x,t) = \chi_2(x,t) e^{-ib t + x \mu + t \Omega},
\end{equation}
where $\mu,\Omega \in \mathbb{C}$ are spectral parameters 
and $\chi = (\chi_1,\chi_2)^T$ satisfies the following spectral problems:
\begin{equation}
\label{lin-alg-1-mod}
\chi_x + \mu \chi = \left(\begin{array}{cc} \lambda & \phi \\ -\bar{\phi} & -\lambda \end{array} \right) \chi, 
\end{equation}
and
\begin{equation}
\label{lin-alg-2-mod}
\chi_t + \Omega \chi = i \left(\begin{array}{cc}
\lambda^2 + \frac{1}{2} |\phi|^2 - b & \frac{1}{2} \frac{d\phi}{dx} + \lambda \phi \\
\frac{1}{2} \frac{d \bar{\phi}}{dx} - \lambda\bar{\phi} & -\lambda^2 - \frac{1}{2} |\phi|^2 + b \end{array} \right) \chi.
\end{equation}
Parameters $\mu, \Omega \in \mathbb{C}$ are independent of 
$(x,t)$. This follows from the same compatibility of the linear Lax equations (\ref{3.1}) and (\ref{3.2}) if $\psi(x,t)$ in (\ref{double-periodic-mod}) satisfies the NLS equation (\ref{nls}).

By Floquet theorem, spectral parameters $\mu,\Omega \in \mathbb{C}$ are determined from the periodicity conditions $\chi(x+L,t) = \chi(x,t+T) = \chi(x,t)$ in terms of the spectral parameter $\lambda$. 
We distinguish between the space coordinate $x$ and the time coordinate $t$ in order to consider stability of the double-periodic waves (\ref{double-periodic-mod}) in the time evolution of the NLS equation (\ref{nls}). 

{\em The Lax spectrum} is defined by the condition that $\lambda$ belongs to an admissible set for which the solution (\ref{lax-eq-mod}) is bounded in $x$. Hence $\mu = i \theta$ with real $\theta$ in $\left[-\frac{\pi}{L},\frac{\pi}{L} \right]$ 
and $\lambda$ is computed from the spectral problem (\ref{lin-alg-1-mod}) with $\chi(x+L,t) = \chi(x,t)$ for every $t \in \mathbb{R}$. 

With $\lambda$ defined in the Lax spectrum, the spectral problem (\ref{lin-alg-2-mod}) can be solved for the spectral parameter $\Omega$ under the condition that $\chi(x,t+T) = \chi(x,t)$ for every $x \in \mathbb{R}$. 
The corresponding solution to the linear system (\ref{lin-alg-1-mod}) and (\ref{lin-alg-2-mod}) generates the solution $v(x,t)$ 
of the linearized system (\ref{nls-lin}) with $u \equiv \phi$ and $c = 0$ by means of the transformation formulas (\ref{w-eigen}) and (\ref{eig-stability}). 
Spectral parameter $\Omega$ is uniquely defined in the fundamental 
strip ${\rm Im}(\Omega) \in [-\frac{\pi}{T},\frac{\pi}{T}]$, 
while ${\rm Re}(\Omega)$ determines the instability rate ${\rm Re}(\Lambda)$ by $\Lambda = 2 \Omega$. 

If ${\rm Re}(\Lambda) > 0$ for $\lambda$ in the Lax spectrum, the double-periodic wave (\ref{double-periodic-mod}) is
called {\em spectrally unstable}. The amplitude-normalized double-periodic waves are taken by using the scaling transformation (\ref{scaling-transform}). 
We observe again that the unstable spectrum with ${\rm Re}(\Lambda) > 0$ is related with $\lambda$ in the Lax spectrum outside the imaginary axis. 

\begin{figure}[htp!]
	\centering
	\includegraphics[width=8cm,height=5.5cm]{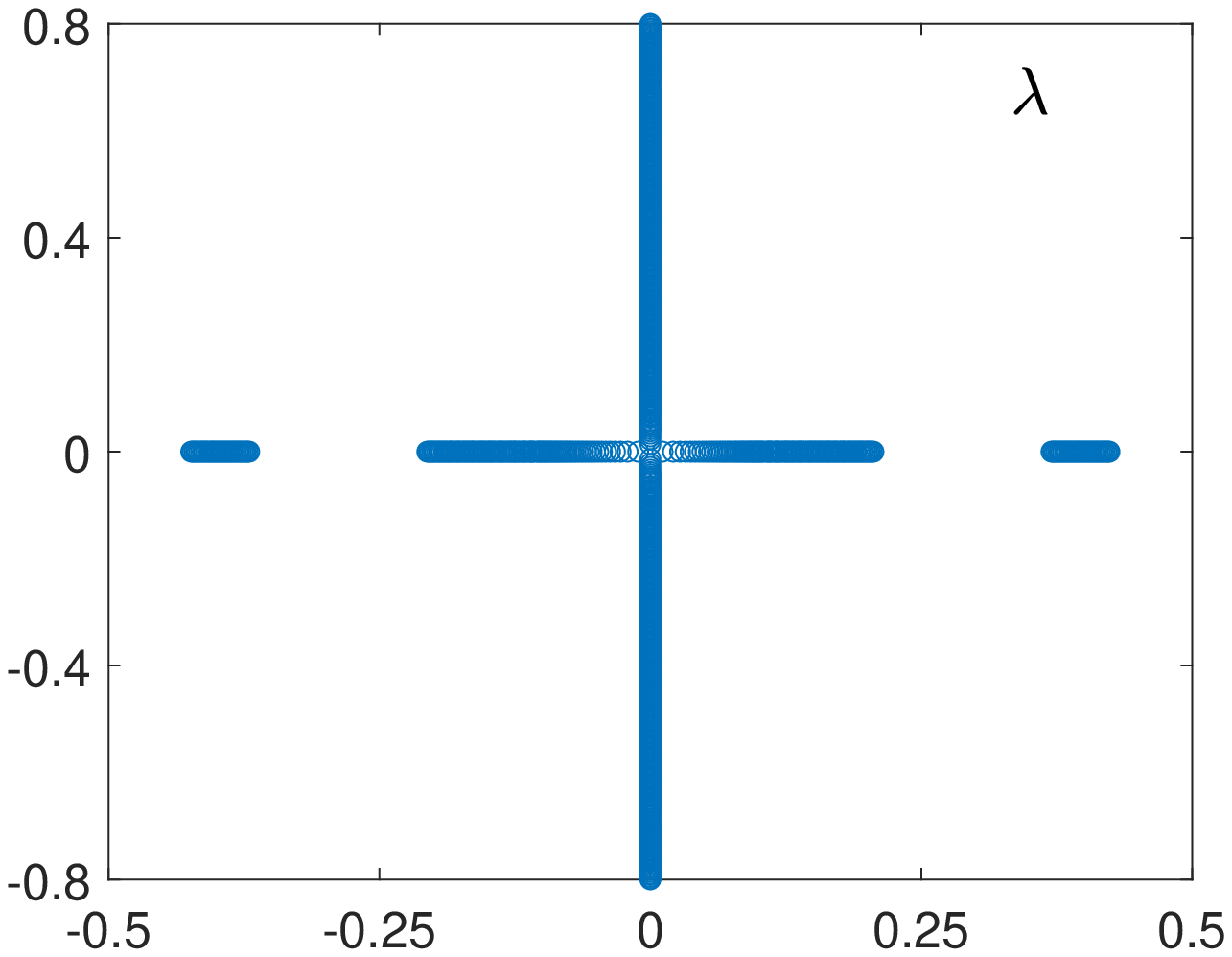}
	\includegraphics[width=8cm,height=5.5cm]{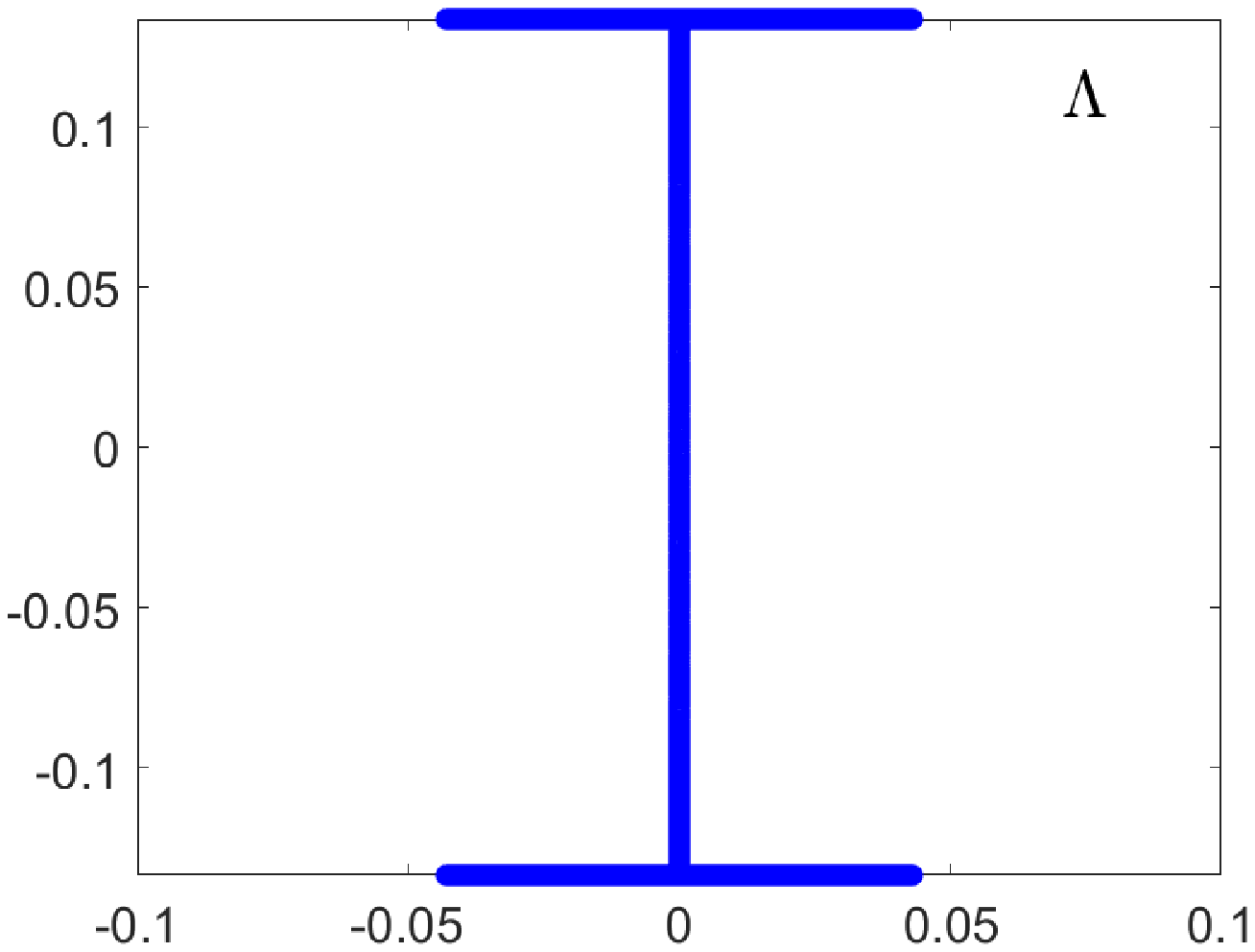} \\
	\includegraphics[width=8cm,height=5.5cm]{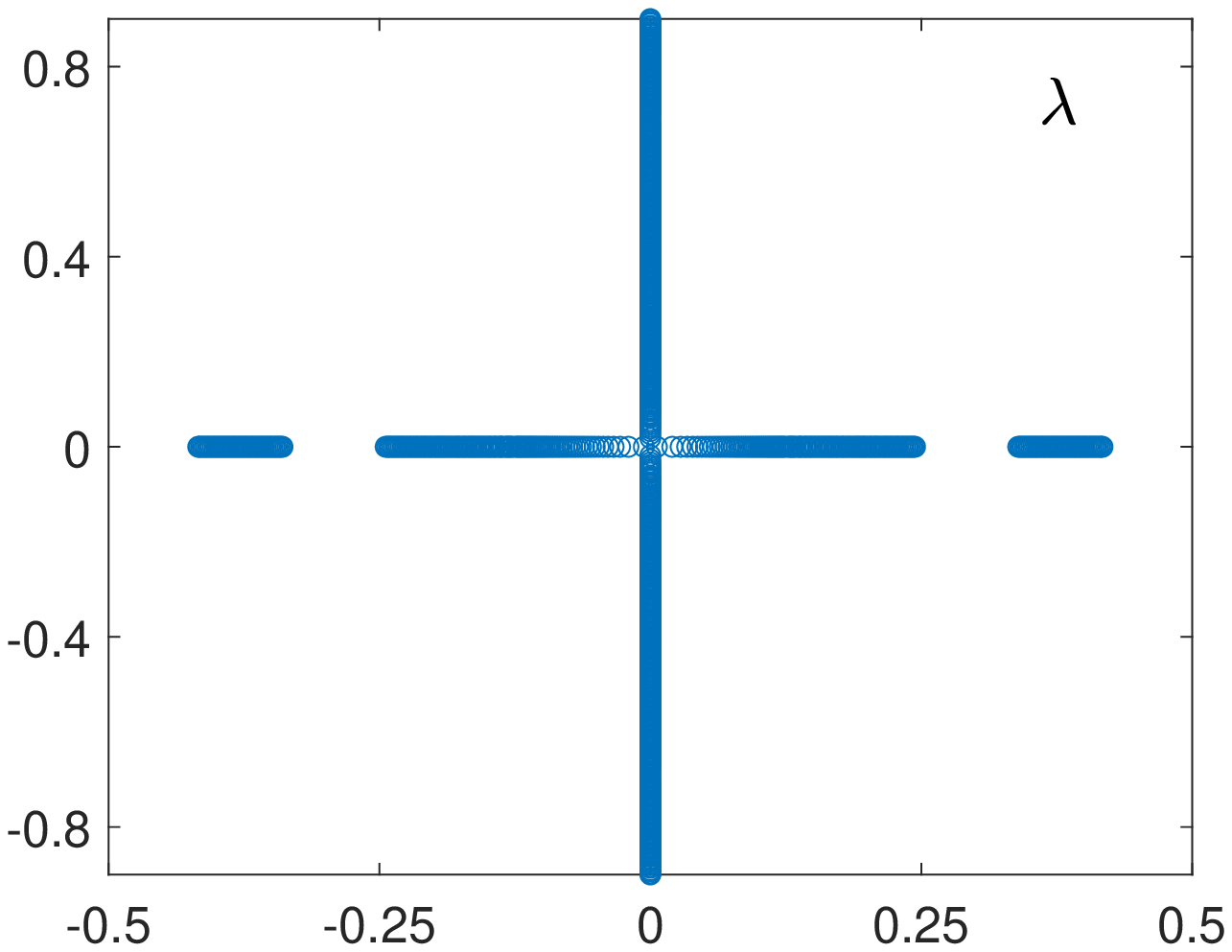}
	\includegraphics[width=8cm,height=5.5cm]{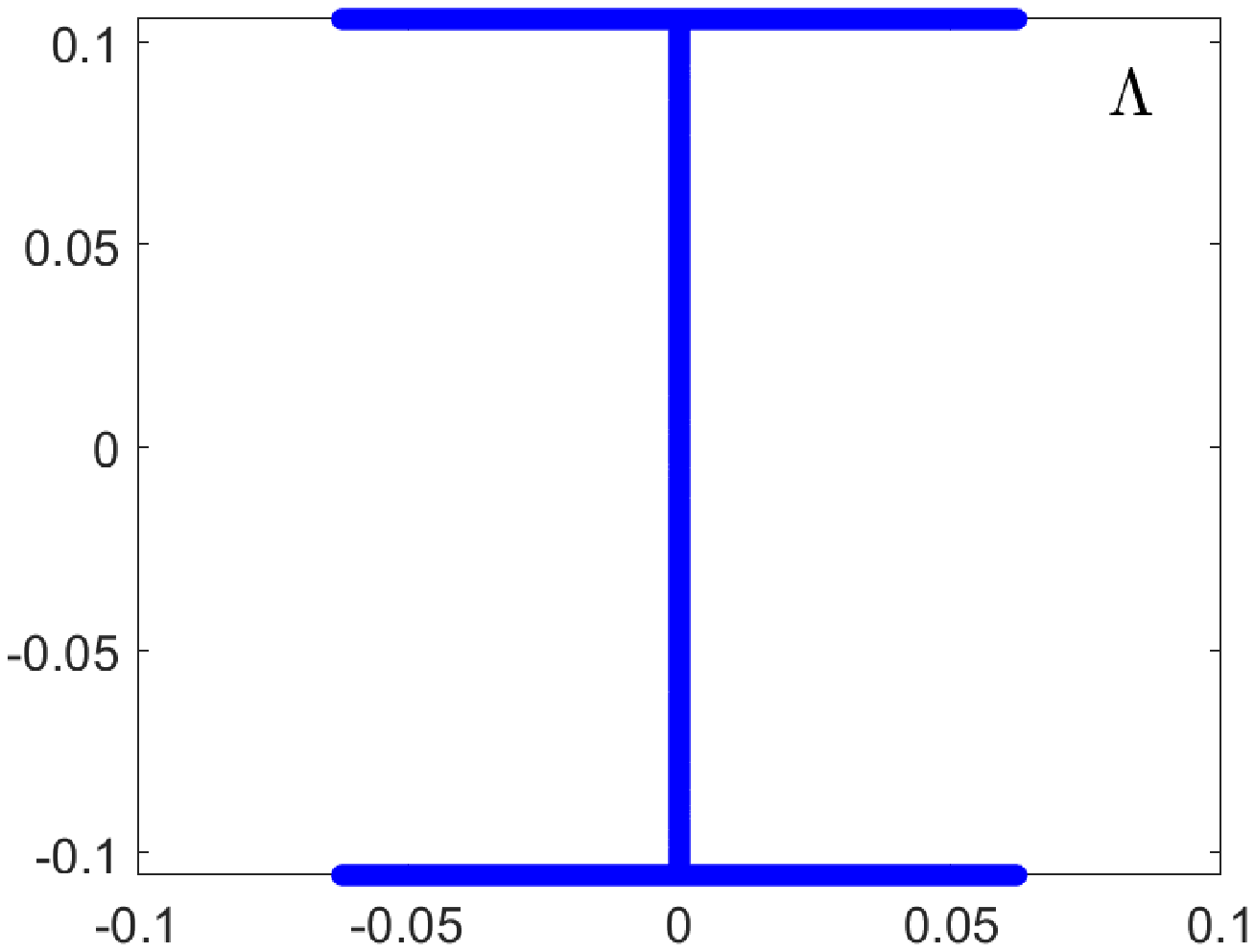}
	\caption{Lax spectrum on the $\lambda$-plane (left) and stability spectrum on the $\Lambda$-plane (right) for the amplitude-normalized double-periodic wave (\ref{solB}) with $k=0.85$ (top) and $k = 0.95$ (bottom).}
	\label{f3}
\end{figure}

\begin{figure}[htp!]
	\centering
	\includegraphics[width=8cm,height=6.5cm]{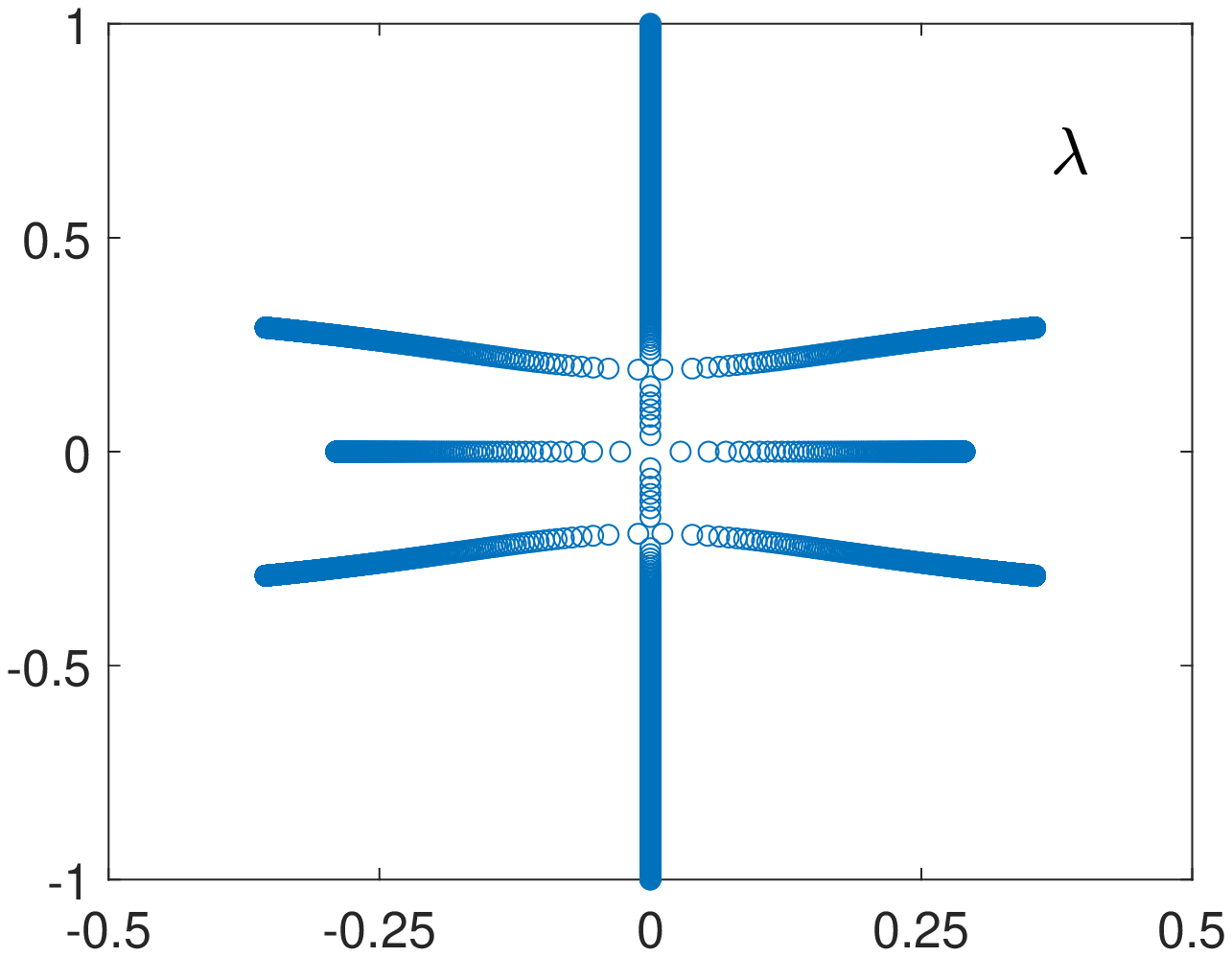}
	\includegraphics[width=8cm,height=6.5cm]{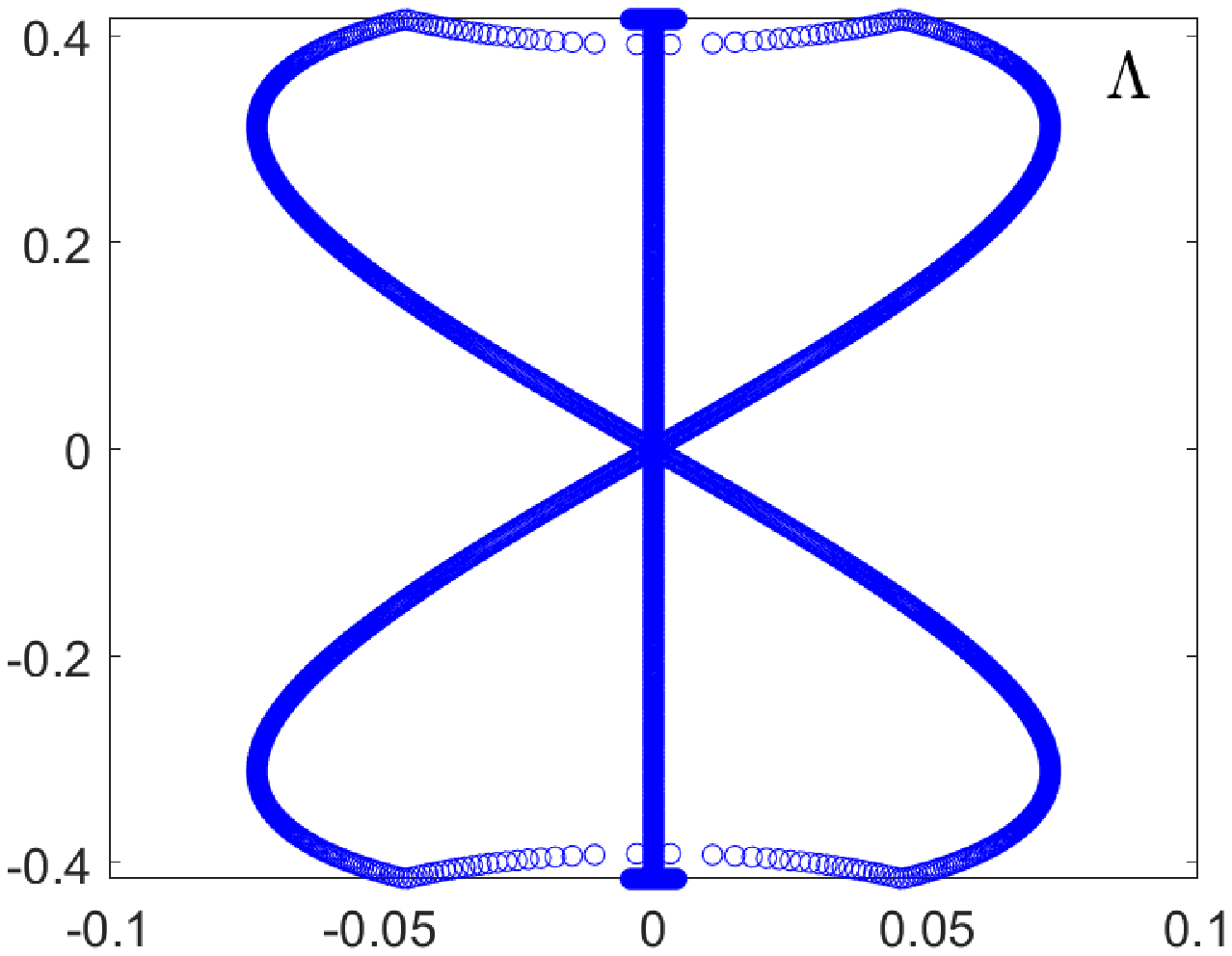} \\
	\includegraphics[width=8cm,height=6.5cm]{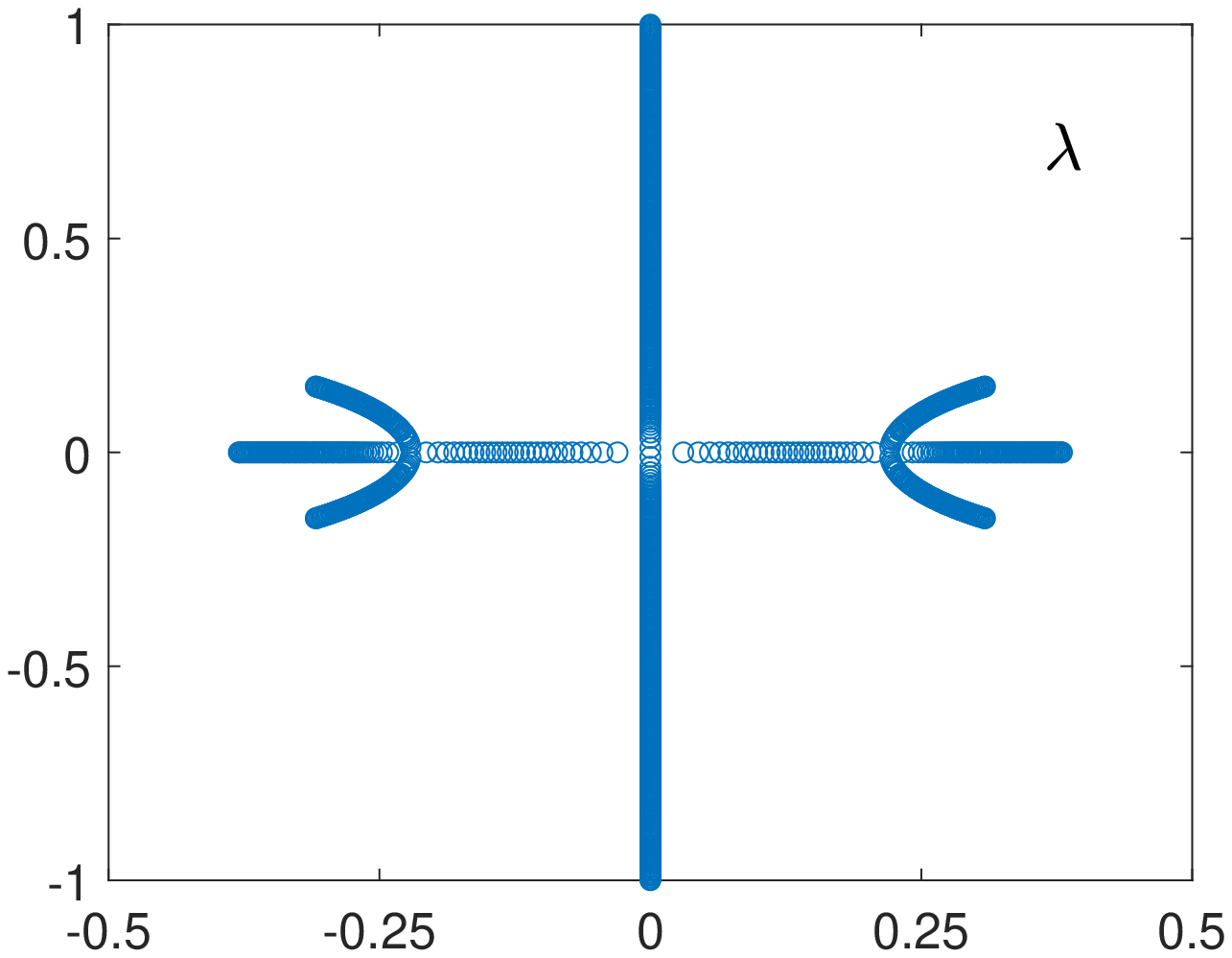}
	\includegraphics[width=8cm,height=6.5cm]{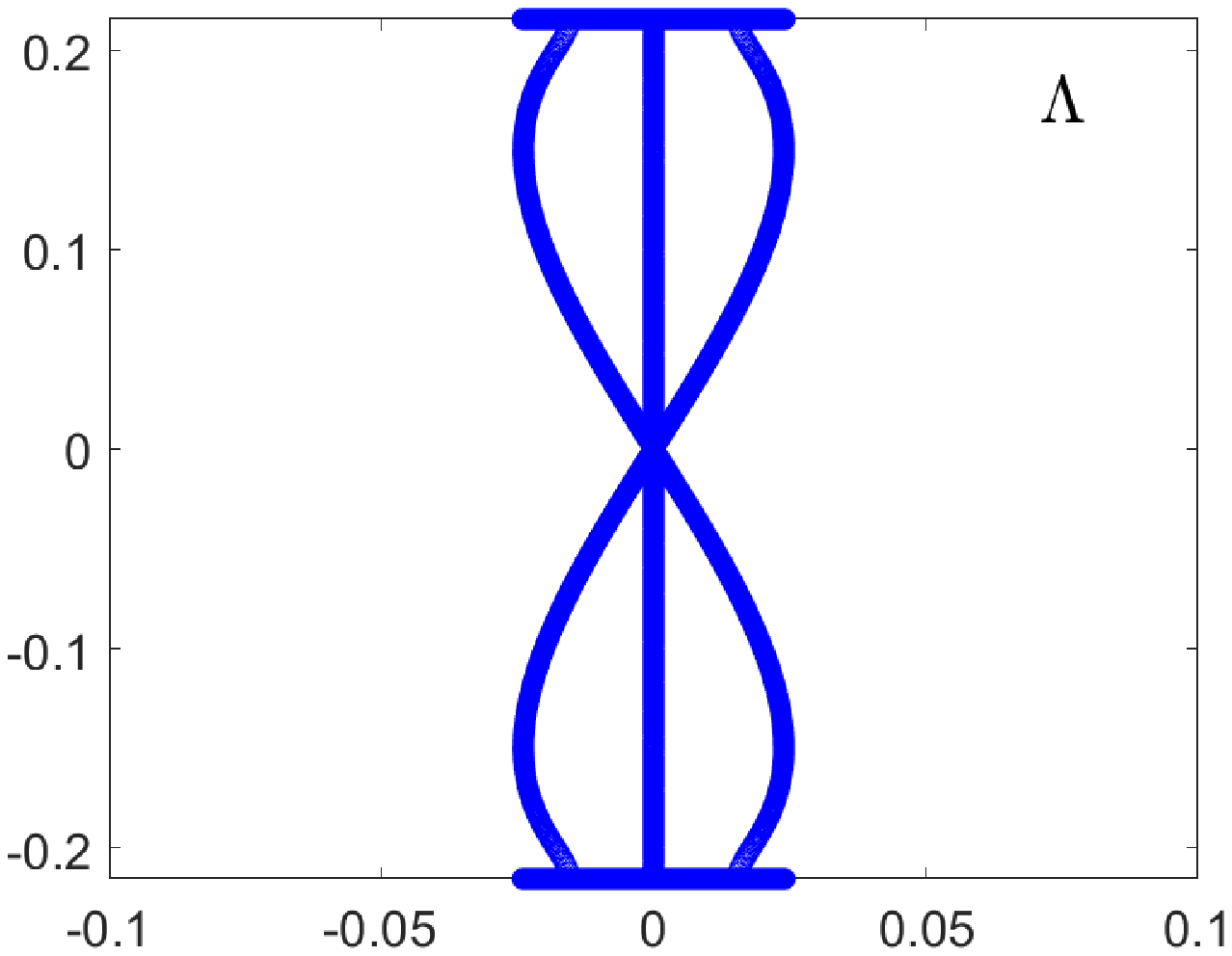} \\
	\includegraphics[width=8cm,height=6.5cm]{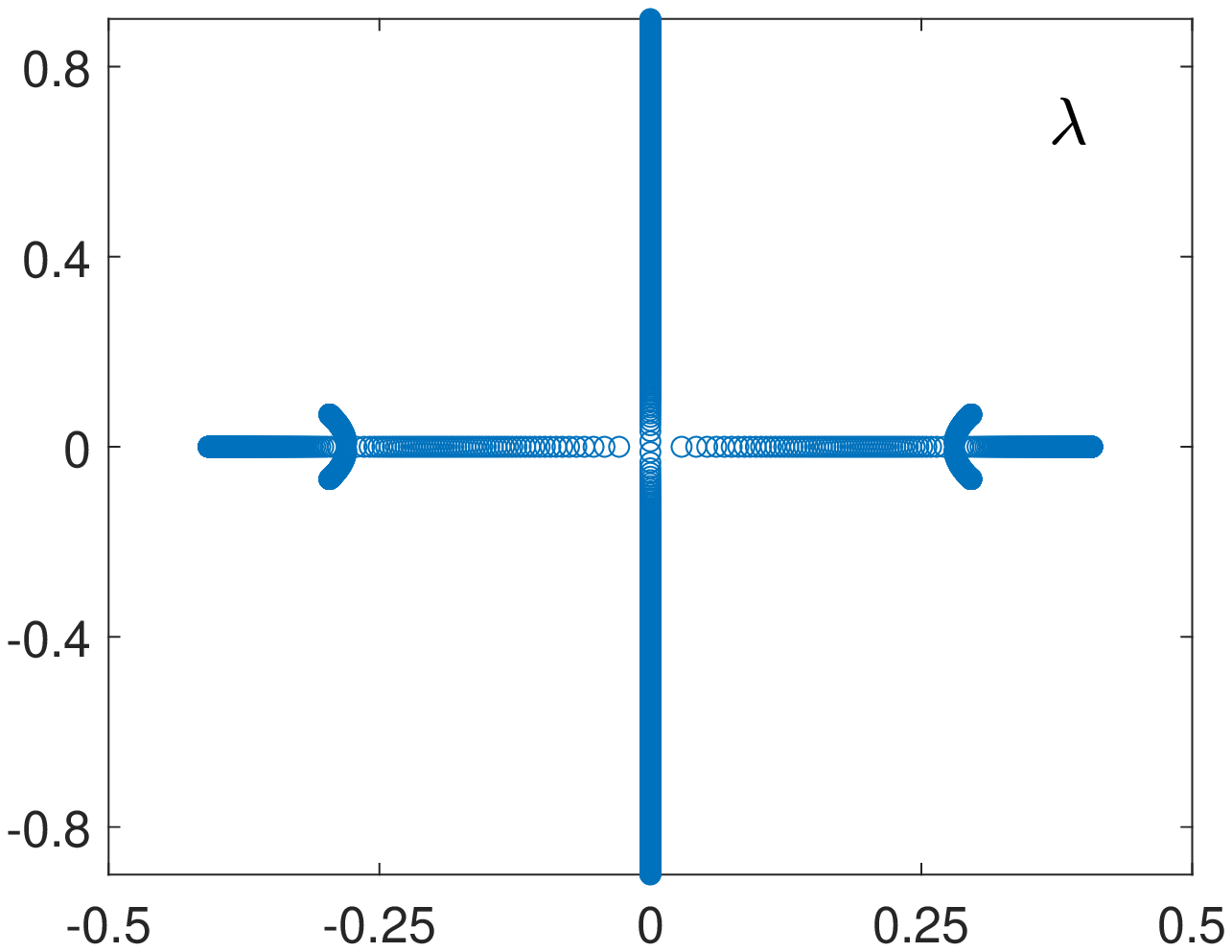}
	\includegraphics[width=8cm,height=6.5cm]{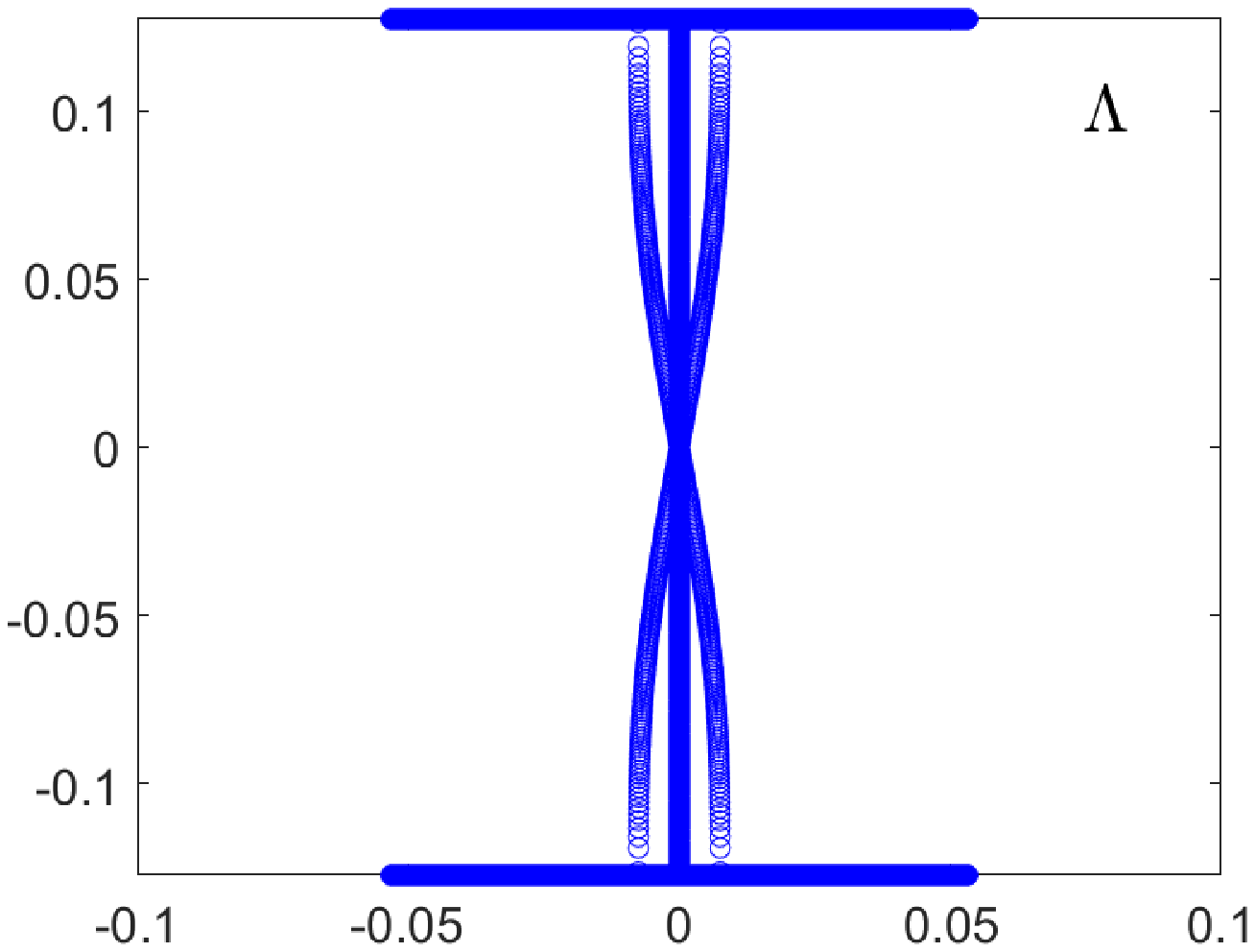}
	\caption{The same as Figure \ref{f3} but for the amplitude-normalized double-periodic wave (\ref{solA}) with $k=0.3$ (top), $k = 0.6$ (middle), and $k = 0.9$ (bottom).}
	\label{f4}
\end{figure}

For the amplitude-normalized double-periodic wave (\ref{solB}) with $k = 0.85$ (top) and $k = 0.95$ (bottom), Fig. \ref{f3} shows the Lax spectrum of the spectral problem (\ref{lin-alg-1-mod}) with $\mu = i \theta$ and $\theta \in [-\frac{\pi}{L},\frac{\pi}{L}]$ on the $\lambda$-plane (left) and the stability spectrum on the $\Lambda$-plane (right).  
The unstable spectrum is located 
at the boundary ${\rm Im}(\Lambda) = \pm \frac{2 \pi}{T}$ of the strip for every $k \in (0,1)$. 
The double-periodic wave (\ref{solB}) is spectrally unstable.

Fig. \ref{f4} shows the same as Fig. \ref{f3} but for the amplitude-normalized double-periodic 
wave (\ref{solA}) with $k = 0.3$ (top), $k = 0.6$ (middle), and $k = 0.9$ (bottom). The Lax spectrum on the $\lambda$-plane has three bands, two of which are connected either across the imaginary axis (top) or across the real axis (middle and bottom), the third band is located on the real axis. The unstable spectrum on the $\Lambda$-plane 
includes the figure-eight band and bands located near the boundary ${\rm Im}(\Lambda) = \pm \frac{2 \pi}{T}$. As $k \to 1$, the figure-eight band becomes very thin and the stability spectrum looks similar 
to the one on Fig. \ref{f3} because both the double-periodic solutions approach the Akhmediev breather (\ref{Akh-breather}).

\begin{figure}[htp!]
	\centering
	\includegraphics[width=8cm,height=5cm]{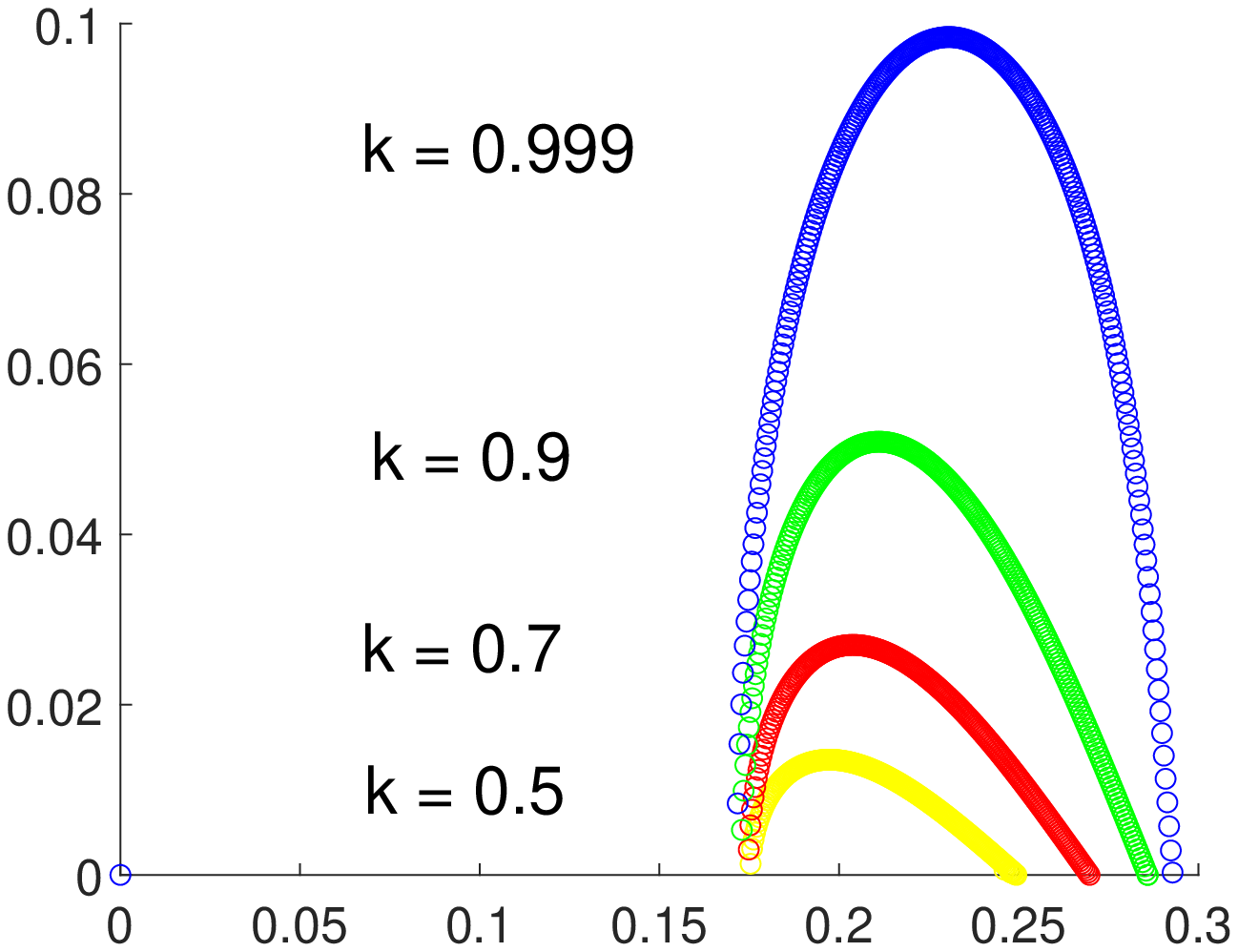}
	\includegraphics[width=8cm,height=5cm]{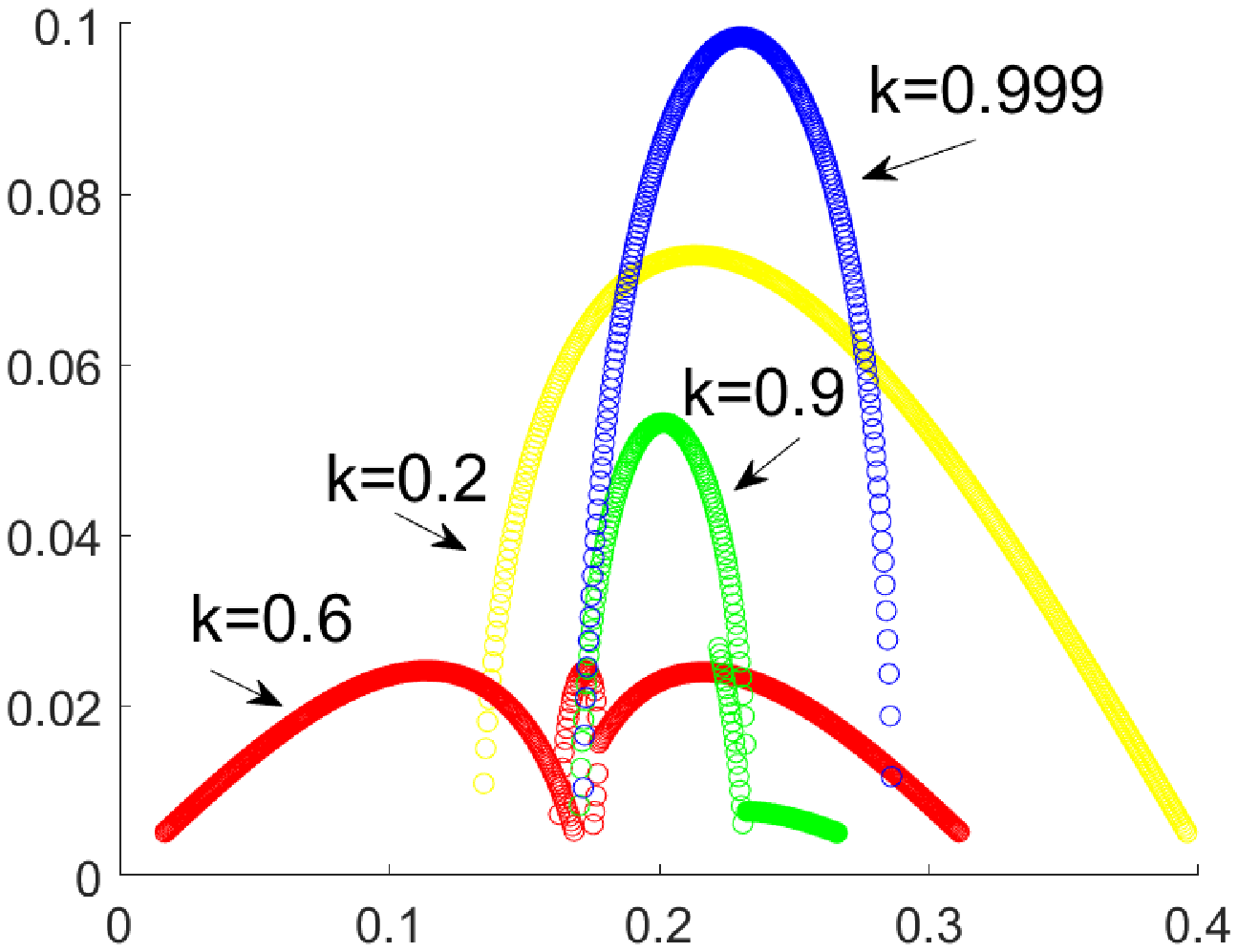}
	\caption{Instability rate ${\rm Re}(\Lambda)$ versus the Floquet parameter $\theta$ for the amplitude-normalized double-periodic wave (\ref{solB}) (left) and the double-periodic wave (\ref{solA})  (right). The values of $k$ are given in the plots.}
	\label{f5}
\end{figure}

Figure \ref{f5} compares the instability rates for different double-periodic waves of the same unit amplitude. ${\rm Re}(\Lambda)$ is plotted versus the Floquet parameter $\theta$ in $[0,\frac{\pi}{L}]$, where $\mu = i \theta$ is defined in (\ref{lin-alg-1-mod}).

For the amplitude-normalized double-periodic wave (\ref{solB}) (left), the instability rate is maximal as $k \to 1$, that is, at the 
Akhmediev breather (\ref{Akh-breather}). The unstable band starts with the same cut-off value of $\theta$ and extends all the way to $\theta = \frac{\pi}{L}$. When $k \to 0$, the instability rates quickly decrease 
as the double-periodic wave approaches the NLS soliton (\ref{NLS-soliton}).

For the amplitude-normalized double-periodic wave (\ref{solA}) (right), the instability rate is large in the limit $k \to 0$, when the double-periodic wave is close to the particular ${\rm cn}$-periodic wave (\ref{NLS-wave}). Then, the rates descrease when $k$ is increased, however, the rates increase again and reach the maximal values as $k \to 1$ when the double-periodic wave approaches the Akhmediev breather (\ref{Akh-breather}).

\section{Conclusion}

We have computed the instability rates for the double-periodic waves of the NLS equation. By using the Lax pair of linear equations, we obtain the Lax spectrum with the Floquet theory in the spatial coordinate at fixed $t$ and the stability spectrum with the Floquet theory in the temporal coordinate at fixed $x$. This separation of variables is computationally simpler than solving the full two-dimensional system of linearized NLS equations on the double-periodic solutions. 

As the main outcome of the method, we have shown instability of the double-periodic solutions and have computed their instability rates, which are generally smaller compared to those for the  standing periodic waves. 

The concept can be extended to other double-periodic solutions of the NLS equation which satisfy the higher-order Lax--Novikov equations. Unfortunately, the other double-periodic solutions are only available 
in Riemann theta functions of genus $d \geq 2$, and for practical computations, 
one needs to construct such double-periodic solutions numerically, similarly to what was done in \cite{Tovbis1}. This task is opened for further work.

\vspace{0.25cm}

{\bf Acknowledgement:} This work was supported in part by the National
Natural Science Foundation of China (No. 11971103).

\end{document}